\documentclass[a4paper,11pt]{article}

\usepackage[utf8]{inputenc}
\usepackage{a4wide}
\usepackage{tabularx} 
\usepackage{multirow}
\usepackage{authblk}
\usepackage{color,xcolor}
\usepackage{amsmath}  
\usepackage{amssymb}
\usepackage{graphicx} 
\usepackage{subcaption} %
\usepackage{cite} 
\usepackage[final]{hyperref} 
\usepackage{booktabs}
\def\linkcolor{cyan!70!black}
\hypersetup{
	colorlinks=true,       
	linkcolor=\linkcolor,        
	citecolor=\linkcolor,        
	filecolor=\linkcolor,     
	urlcolor=\linkcolor         
}
\usepackage{blindtext}
\usepackage{cleveref}
\usepackage{appendix}

\definecolor{orange}{rgb}{1,0.5,0}

\newcommand{\ord}[1]{{\mathcal{O}(#1)}}

\newcommand{\revision}[1]{#1}

\renewcommand{\thefootnote}{\fnsymbol{footnote}}

\begin{document}

\mbox{}\vspace{1cm}

\begin{center}

{\LARGE \bf \boldmath Phenomenology of Non-Abelian Gauge and\\[2mm] Goldstone Bosons in a U(2) Flavor Model}

        \vspace{1cm}

{\Large Lorenzo Calibbi and Jiangyi Yi}\,\footnote{Corresponding author, {\it e-mail}\,: \href{mailto:yijiangyi@nankai.edu.cn}{yijiangyi@nankai.edu.cn}.}

        \vspace{0.5cm}

{\emph{\large School of Physics, Nankai University, Tianjin 300071, China}}

\end{center}
\vspace*{1cm}

\begin{abstract}
We investigate the phenomenological implications of the bosons associated with the $SU(2)_F$ subgroup in a simple and realistic $U(2)_F$ flavor model. While the Nambu-Goldstone boson of the $U(1)_F$ factor behaves as a standard QCD axion (an axiflavon) with suppressed flavor-violating couplings, the three degrees of freedom from $SU(2)_F$ have not been studied before. This work focuses on these states, considering both the case where $SU(2)_F$ is a global symmetry, yielding pseudo-Nambu-Goldstone bosons (PNGBs), and the case where it is a gauge symmetry with a potentially small coupling, yielding a triplet of (possibly) light gauge bosons. In both scenarios, these new bosons naturally feature unsuppressed flavor-violating couplings to Standard Model fermions in the mass basis. We derive the resulting predictions for flavor-changing neutral currents and lepton flavor violation, including exotic decays of mesons and leptons. Our analysis shows that processes like $K \to \pi X$ and $\mu \to e X$ place the most stringent constraints, probing the flavor symmetry breaking scale $v_\phi$ up to $10^{11}-10^{12}$~GeV for light bosons, while heavier states are tested in $B$ and $\tau$ decays, as well as by $K-\bar K$ mixing and $\mu\to e \gamma$. We demonstrate that low-energy flavor experiments provide a powerful probe of this framework, capable of testing ultra-high symmetry breaking scales that surpass the limits set by astrophysical observations.
\end{abstract}

\thispagestyle{empty}

\renewcommand{\thefootnote}{\arabic{footnote}}

\newpage


\section{Introduction}
\setcounter{footnote}{0}

The Standard Model (SM) of particle physics provides a theoretically consistent and experimentally successful description of all known particles and interactions.
Nevertheless, a number of fundamental observations about our universe notoriously find no explanation within the SM domain: the origin of neutrino masses, the nature of dark matter, the asymmetry of baryons and antibaryons in the observable universe, the field origin of dark energy and inflation.
In addition to these open questions\,---\,which unambiguously demand an extension of the SM\,---\,there are further issues of a somewhat different nature, which stem from the lack of an organizing principle accounting for what, in the SM, are merely free parameters. Well-known examples include the SM flavor puzzle and the strong CP problem, both of which we address in the present work.

``Flavor puzzle'' in particle physics refers to the challenge of explaining the very existence of three generations of quarks and leptons and the peculiar (mostly) hierarchical patterns of their masses and mixing angles that, within the SM, stem from the Yukawa couplings with the Higgs field\,---\,see e.g.~\cite{Feruglio:2015jfa,Xing:2020ijf,Altmannshofer:2024ykf}.
Arguably, this is not merely a matter of explaining why measured numbers are what they are: in fact, fermion masses and mixing display regularities that seem to suggest the existence of an underlying dynamics. Hence, exploring the flavor puzzle may open a window onto a new physics sector also addressing the other problems mentioned above. 

The ``strong CP problem'' is the fact that, so far, no CP-violating effects stemming from the operator $\mathcal{L} \supset \bar \theta \frac{g_s}{32\pi^2} G\widetilde G$ (where $G$ and $\widetilde G$ are the QCD field strength and its dual, and $g_s$ is the strong coupling) have been observed, implying that the QCD effective angle $\bar \theta$ must have a tiny value, $\bar \theta < 10^{-10}$\,---\,see~\cite{Hook:2018dlk} for an introduction to the subject. Interestingly, one can observe a twofold connection between the strong CP problem and the origin of fermion masses, that is, the flavor puzzle: (i)~a contribution to $\bar\theta$ arises from quark masses (see e.g.~\cite{DiLuzio:2020wdo} for a review); 
(ii)~the Peccei-Quinn (PQ) symmetry that can provide a dynamical solution to the problem~\cite{Peccei:1977hh,Weinberg:1977ma,Wilczek:1977pj} is a subgroup of the large SM global flavor symmetry group $U(3)^5$~\cite{DAmbrosio:2002vsn}.
Furthermore, the interesting possibility that another, flavor non-universal, subgroup of $U(3)^5$ could act as a PQ symmetry, leading to a flavor-violating axion, was suggested by Wilczek long ago~\cite{Wilczek:1982rv}\,---\,see also \cite{Davidson:1981zd,Reiss:1982sq,Davidson:1983fy,Davidson:1983fe,Chang:1987hz,Berezhiani:1990wn,Berezhiani:1990jj,Feng:1997tn,Albrecht:2010xh} for related early discussions on Nambu-Goldstone bosons from flavor symmetries, including non-Abelian examples.

One promising framework for addressing the above issues involves the spontaneous breaking of (a subgroup of) the mentioned flavor symmetry. Being the SM fermions charged under such symmetry, the Yukawa couplings are typically forbidden in the symmetric limit. The spontaneous breaking of the flavor symmetry then occurs through the vacuum expectation values (VEVs) of new scalar fields\,---\,the so-called ``flavons''\,---\,leading to Yukawa interactions that are suppressed because arising from higher-dimensional operators~\cite{Froggatt:1978nt,Leurer:1992wg,Leurer:1993gy}.  

Interestingly, it has been shown that this framework can simultaneously address the strong CP problem~\cite{Ema:2016ops,Calibbi:2016hwq}, even within its simplest realization based on the flavor non-universal global $U(1)$ symmetry originally proposed by Froggatt and Nielsen (FN) in~\cite{Froggatt:1978nt}, hence minimally realizing Wilczek's idea mentioned above~\cite{Wilczek:1982rv}. Indeed, the FN $U(1)$ charge assignment of SM fermions is typically anomalous under QCD. As a consequence, the resulting pseudo Nambu-Goldstone boson (PNGB)\,---\,the ``axiflavon''\,---\,can solve the strong CP problem, dynamically driving the QCD vacuum to a CP-conserving minimum, acquires mass from the QCD anomaly, and can account for the observed cold dark matter abundance through the misalignment mechanism, just like a QCD axion~\cite{Ema:2016ops,Calibbi:2016hwq}. In addition, its existence can be constrained (or discovered!) by searches for exotic flavor-violating decays such as $K\to \pi X$, with $X$ being a light invisible boson~\cite{NA62:2021zjw,NA62:2025upx}. 

The simplicity and attractive features of flavor models based on a FN $U(1)$ symmetry explain why they are the object of persistent interest in the literature\,---\,see e.g.~\cite{Smolkovic:2019jow,Fedele:2020fvh,Nishimura:2020nre,Cornella:2023zme,Nishimura:2024apb,Blasi:2024vew,Ibe:2024cvi,Cornella:2024jaw} for recent examples.
Nevertheless, this class of models also has some unsatisfactory aspects. In particular:
(i)~the large number of free parameters\,---\,especially the $\ord{1}$ coefficients that appear in each entry of the Yukawa matrices\,---\,limits its predictive power;
(ii)~typical UV completions require the introduction of a great number of fields, as systematically discussed in~\cite{Calibbi:2012yj,Calibbi:2012at};
(iii)~an Abelian symmetry is unable to provide a rationale for the existence of three fermion generations.

All of these drawbacks can be addressed, or at least ameliorated, within models with a non-Abelian flavor symmetry. 
The simplest example\,---\,and perhaps the best motivated by the observed pattern of fermion masses\,---\,is provided by $U(2)$.
Third-generation fermions (top, bottom and tau) are significantly heavier than the corresponding first two generations of quarks and leptons. Hence, the latter can be considered in first approximation massless, as they would be if belonging to doublets of an unbroken symmetry, under which the third generations are singlets. As a result, the non-Abelian $U(2)$ flavor symmetry naturally emerges as a promising candidate to describe the flavor structure observed in the data~\cite{Barbieri:1995uv,Barbieri:1997tu,Roberts:2001zy,Dudas:2013pja,Falkowski:2015zwa,Linster:2018avp,Barbieri:2019zdz,Darme:2023nsy,Greljo:2023bix}.
A simple and realistic flavour model based on   $U(2)_F \simeq SU(2)_F \times U(1)_F$  has been proposed in \cite{Linster:2018avp}, with charge assignments compatible with the unification of quarks and leptons within $SU(5)$. 
In this framework, two flavons are introduced, an $SU(2)_F$ doublet and a singlet, hence the $U(2)_F$ symmetry breaking is controlled by two small parameters related to the VEVs of such fields. The resulting operators provide an excellent fit to all SM flavor observables, including neutrino masses and mixing angles~\cite{Linster:2018avp,Linster:2020fww,Calibbi:2020jvd,Giarnetti:2025idu}.


In this context, the PNGB associated with $U(1)_F \subset U(2)_F$ behaves like a QCD axion, that is, it is an axiflavon, albeit with a different phenomenology compared to the FN $U(1)$ case. In particular, the $U(2)_F$ axiflavon features suppressed flavor-violating interactions preventing its discovery in meson or lepton decays~\cite{Linster:2018avp}. This makes it difficult to discriminate this appealing scenario from other QCD axion models based on the axiflavon phenomenology alone, unless the charge assignment is modified in a way incompatible with $SU(5)$ unification~\cite{Calibbi:2020jvd}.

In the original work~\cite{Linster:2018avp}, the three Nambu-Goldstone bosons of the remaining $SU(2)_F$ are not discussed, arguably following the assumption that the symmetry is local and, thus, these fields constitute the longitudinal components of a triplet of gauge bosons of little phenomenological relevance because superheavy.
While nothing is inconsistent with such an approach, the present work focuses precisely on the phenomenology of such fields, in search of distinctive observable phenomena that might provide experimental tests of the $U(2)_F$ model.

The following discussion considers both the cases of a global and a gauged $SU(2)_F$, both scenarios being consistent because the symmetry is anomaly free. In either case, we leave $U(1)_F$ global in order to address the strong CP problem and have a dark matter candidate through the axiflavon.
In the global case, we assume that the three additional bosons acquire mass through a (small) explicit symmetry breaking term, thereby becoming PNGBs. 
In the local $SU(2)_F$ scenario, we allow the resulting gauge bosons to be much lighter than the symmetry breaking scale\,---\,that is, we entertain the possibility that the $SU(2)_F$ gauge coupling is small. In either case, the new light bosons feature flavor-violating couplings with fermions (in the fermion mass basis) by construction. Hence, they mediate flavor-changing neutral currents, which lead to transitions between different generations of quarks and leptons or exotic meson and lepton decays into these new light states. We show that these processes might leave measurable imprints in low-energy observables, while evading astrophysical constraints (from star cooling and supernova explosions) such that they could be discovered at running and future flavor experiments.

The rest of the paper is organized as follows.  In Section~\ref{sec:model}, we review the framework and discuss the Nambu-Goldstone bosons arising from the spontaneous breaking of the $U(2)_F$ flavor symmetry. In Section~\ref{sec:local}, we discuss the case where $SU(2)_F$ is a local symmetry, entailing the existence of a triplet of flavored gauge bosons $W^\prime$.
Section~\ref{sec:global} is instead devoted to introducing the PNGB triplet $\pi^\prime$ in the case of a global $SU(2)_F$. The phenomenological implications of both cases are presented in Section~\ref{sec:pheno}. Finally, we summarize our findings and draw our conclusions in Section~\ref{sec:concl}. Useful formulae and experimental input are collected in the appendices.


\section{Set-up of the model}
\label{sec:model}

\begin{table}[t!]
    \centering
\begin{tabular}{l*{13}{c}}
\toprule
 & $U_a$ & $D_a$ & $Q_a$ & $U_3$ & $D_3$ & $Q_3$ & $E_a$ & $L_a$ & $E_3$ & $L_3$ & $H$ & $\phi_a$ & $\chi$ \\ \midrule
\({SU}(2)_F\) & $\mathbf{2}$ & $\mathbf{2}$ & $\mathbf{2}$ & $\mathbf{1}$ & $\mathbf{1}$ & $\mathbf{1}$ & $\mathbf{2}$ & $\mathbf{2}$ & $\mathbf{1}$ & $\mathbf{1}$ & $\mathbf{1}$ & $\mathbf{2}$ & $\mathbf{1}$ \\ \midrule
\({U}(1)_F\) & 1 & 1 & 1 & 0 & 1 & 0 & 1 & 1 & 0 & -1 & 0 & -1 & -1 \\ \bottomrule
\end{tabular}
    \caption{Transformation properties of the fields under $U(2)_{F} \simeq SU(2)_F \times U(1)_F$. $Q$ and $L$ are respectively left-handed (LH) quark and lepton doublets. $U$, $D$, $E$ are respectively right-handed (RH) up-type quark, down-type quark and lepton singlets. The subscripts denote flavor indices and, in particular, $a=1,2$ is the index of the fundamental representation of $SU(2)_F$. $H$ denotes the SM Higgs field, while $\phi_a$ and $\chi$ are the flavon fields.}
    \label{tab:charges}
\end{table}

In this section, we review the basic framework of the $U(2)_F$ flavor model~\cite{Linster:2018avp} and introduce the Nambu-Goldstone bosons associated with the spontaneous breaking of the flavor symmetry. 
The transformation properties of the SM fermions under $U(2)_F \simeq SU(2)_F \times U(1)_F$ are displayed in Table~\ref{tab:charges}.
The first two generations of fermions transform as doublets under $SU(2)_F$, while the third generation is an $SU(2)_F$ singlet. Hence, we adopt the notation $Q_a = ( Q_1 \ Q_2 )^T$ to represent the first and second generation LH quarks, and similarly for the LH leptons. The Higgs field is a singlet under $SU(2)_F$ and is $U(1)_F$-neutral. 

In the original $U(2)_F$ model~\cite{Linster:2018avp}, the SM fermions were assigned $SU(2)_F \times U(1)_F$ quantum numbers consistent with an $SU(5)$ grand unified (GUT) structure and neutrinos were assumed to be Dirac particles in order to achieve a consistent fit of neutrino masses and mixing. Recently, a suitable charge assignment for RH neutrinos has been shown to lead to realistic Majorana neutrino masses via the seesaw mechanism~\cite{Giarnetti:2025idu}.\footnote{For a general discussion of the neutrino sector within the $U(2)_F$ framework, see~\cite{Linster:2020fww}.}  Here, we adopt the modified charge assignment introduced in~\cite{Calibbi:2020jvd}, where the $U(1)_F$ charge of $L_3$ is $-1$ instead of $+1$, which allows us to write a realistic Majorona neutrino mass matrix without the need to specify the UV completion of the Weinberg operator. However, we note that this choice has a very mild impact on the results presented in the remainder of the paper.

 In order to break the flavor symmetry, two scalar fields (the ``flavons'' $\chi$ and $\phi$) are introduced, with $\chi$ being an $SU(2)_F$ singlet and $\phi$ an $SU(2)_F$ doublet, both carrying $U(1)_F$ charge $-1$ (see Table~\ref{tab:charges}). The VEVs of these fields are
\begin{align}
\left\langle \phi \right\rangle = ~\frac{v_{\phi}} {\sqrt{2}} \begin{pmatrix} 1\\ 0  \end{pmatrix} \equiv \begin{pmatrix} \varepsilon_\phi \Lambda\\ 0  \end{pmatrix} ,\quad \left\langle \chi \right\rangle = ~\frac{v_{\chi}}{\sqrt{2}} \equiv \varepsilon_\chi \Lambda\,,
\label{eq:vevs}
\end{align}
where we defined two expansion parameters $\varepsilon_{\phi} \approx \varepsilon_{\chi} \ll 1$ and $\Lambda$ is a cut-off scale associated with the UV completion of the theory.\footnote{Examples of possible UV completions, comprising heavy scalar and/or vectorlike fermion fields, can be systematically built along the lines of Refs.~\cite{Calibbi:2012yj,Calibbi:2012at}.}

By means of a suitable number of flavon insertions, one can write $U(2)_F$-invariant non-renormalizable operators leading to Yukawa interactions that read (at leading order in $\varepsilon_{\phi,\chi}$)
\begin{align}
\label{eq:Lu}
\mathcal{L}_{u}
=~& \frac{\lambda_{11}^{u}}{\Lambda^{6}}\chi^{4}\left( {\phi_{a}^{*}Q_{a}} \right)\left( {\phi_{b}^{*}U_{b}} \right)H + \frac{\lambda_{12}^{u}}{\Lambda^{2}}\chi^{2}\epsilon_{ab}Q_{a}U_{b}H + \frac{\lambda_{13}^{u}}{\Lambda^{3}}\chi^{2}\left( {\phi_{a}^{*}Q_{a}} \right)U_{3}H + \nonumber \\
& \frac{\lambda_{22}^{u}}{\Lambda^{2}}\left( {\epsilon_{ab}\phi_{a}Q_{b}} \right)\left( {\epsilon_{cd}\phi_{c}U_{d}} \right)H + \frac{\lambda_{23}^{u}}{\Lambda}\left( {\epsilon_{ab}\phi_{a}Q_{b}} \right)U_{3}H + \frac{\lambda_{31}^{u}}{\Lambda^{3}}\chi^{2}Q_{3}\left( {\phi_{a}^{*}U_{a}} \right)H + \nonumber \\
& \frac{\lambda_{32}^{u}}{\Lambda}Q_{3}\left( {\epsilon_{ab}\phi_{a}U_{b}} \right)H + \lambda_{33}^{u}Q_{3}U_{3}H\, ,  
\end{align}
\begin{align}
\label{eq:Ld}
\mathcal{L}_{d}
= ~& \frac{\lambda_{11}^{d}}{\Lambda^{6}}\chi^{4}\left( \phi_{a}^{*}Q_{a} \right)\left( \phi_{b}^{*}D_{b} \right)\widetilde{H} + \frac{\lambda_{12}^{d}}{\Lambda^{2}}\chi^{2}\epsilon_{ab}Q_{a}D_{b}\widetilde{H} + \frac{\lambda_{13}^{d}}{\Lambda^{4}}\chi^{3}\left( \phi_{a}^{*}Q_{a} \right)D_{3}\widetilde{H} +\nonumber\\
 &\frac{\lambda_{22}^{d}}{\Lambda^{2}}\left( \epsilon_{ab}\phi_{a}Q_{b} \right)\left( \epsilon_{cd}\phi_{c}D_{d} \right)\widetilde{H} + \frac{\lambda_{23}^{d}}{\Lambda^{2}}\chi\left( \epsilon_{ab}\phi_{a}Q_{b} \right)D_{3}\widetilde{H} + \frac{\lambda_{31}^{d}}{\Lambda^{3}}\chi^{2}Q_{3}\left( \phi_{a}^{*}D_{a} \right)\widetilde{H}  + \nonumber\\
& \frac{\lambda_{32}^{d}}{\Lambda}Q_{3}\left( \epsilon_{ab}\phi_{a}D_{b} \right)\widetilde{H} + \frac{\lambda_{33}^{d}}{\Lambda}\chi Q_{3}D_{3}\widetilde{H}\, , 
\end{align}
\begin{align}
\label{eq:Le}
\mathcal{L}_{e} 
=~& \frac{\lambda_{11}^{e}}{\Lambda^{6}}\chi^{4}\left( \phi_{a}^{*}L_{a} \right)\left( \phi_{b}^{*}E_{b} \right)\widetilde{H} + \frac{\lambda_{12}^{e}}{\Lambda^{2}}\chi^{2}\epsilon_{ab}L_{a}E_{b}\widetilde{H} + \frac{\lambda_{13}^{e}}{\Lambda^{3}}\chi^{2}\left( \phi_{a}^{*}L_{a} \right)E_{3}\widetilde{H} +\nonumber\\
& \frac{\lambda_{22}^{e}}{\Lambda^{2}}\left( \epsilon_{ab}\phi_{a}L_{b} \right)\left( \epsilon_{cd}\phi_{c}E_{d} \right)\widetilde{H} + \frac{\lambda_{23}^{e}}{\Lambda}\left( \epsilon_{ab}\phi_{a}L_{b} \right)E_{3}\widetilde{H} + \frac{\lambda_{31}^{e}}{\Lambda^{4}}\chi^{3}L_{3}\left( \phi_{a}^{*}E_{a} \right)\widetilde{H} + \nonumber\\
& \frac{\lambda_{32}^{e}}{\Lambda^{2}}\chi L_{3}\left( \epsilon_{ab}\phi_{a}E_{b} \right)\widetilde{H} + \frac{\lambda_{33}^{e}}{\Lambda}\chi L_{3}E_{3}\widetilde{H}\,,
\end{align}
where $\lambda_{ij}^{f}$ are dimensionless coefficients (stemming from products of the fundamental couplings of the UV-complete theory) that are assumed to be $\ord{1}$, such that, after inserting the flavon VEVs, the Yukawa hierarchies emerge solely from powers of the small parameters $\varepsilon_\phi$ and $\varepsilon_\chi$ defined in Eq.~\eqref{eq:vevs}. 
\revision{In particular, notice that the above operators, hence the Yukawa interactions, do not depend on the absolute scale of the flavon VEVs nor on the cutoff scale $\Lambda$. Thus, these scales cannot be determined in terms of measured fermion masses and mixing but can only be constrained by flavour processes, as discussed in the following sections.
In terms of the expansion parameters and the $\ord{1}$ coefficients, the Yukawa matrices result:}
\begin{equation}\begin{gathered}
Y^u\approx
\begin{pmatrix}
\lambda_{11}^u\varepsilon_\phi^2\varepsilon_\chi^4 & \lambda_{12}^u\varepsilon_\chi^2 & \lambda_{13}^u\varepsilon_\phi\varepsilon_\chi^2 \\
-\lambda_{12}^u\varepsilon_\chi^2 & \lambda_{22}^u\varepsilon_\phi^2 & \lambda_{23}^u\varepsilon_\phi \\
\lambda_{31}^u\varepsilon_\phi\varepsilon_\chi^2 & \lambda_{32}^u\varepsilon_\phi & \lambda_{33}^u
\end{pmatrix},\quad 
Y^d\approx
\begin{pmatrix}
\lambda_{11}^d\varepsilon_\phi^2\varepsilon_\chi^4 & \lambda_{12}^d\varepsilon_\chi^2 & \lambda_{13}^d\varepsilon_\phi\varepsilon_\chi^3 \\
-\lambda_{12}^d\varepsilon_\chi^2 & \lambda_{22}^d\varepsilon_\phi^2 & \lambda_{23}^d\varepsilon_\phi\varepsilon_\chi \\
\lambda_{31}^d\varepsilon_\phi\varepsilon_\chi^2 & \lambda_{32}^d\varepsilon_\phi & \lambda_{33}^d\varepsilon_\chi
\end{pmatrix}, \\
Y^e\approx
\begin{pmatrix}
\lambda_{11}^e\varepsilon_\phi^2\varepsilon_\chi^4 & \lambda_{12}^e\varepsilon_\chi^2 & \lambda_{13}^e\varepsilon_\phi\varepsilon_\chi^2 \\
-\lambda_{12}^e\varepsilon_\chi^2 & \lambda_{22}^e\varepsilon_\phi^2 & \lambda_{23}^e\varepsilon_\phi \\
\lambda_{31}^e\varepsilon_\phi\varepsilon_\chi & \lambda_{32}^e\varepsilon_\phi\varepsilon_\chi & \lambda_{33}^e\varepsilon_\chi
\end{pmatrix}.
\end{gathered}
\label{eq:yukawas}
\end{equation}

As customary, these matrices can be diagonalized by bi-unitary transformations corresponding to fermion field rotations in the flavor space:
\begin{align}
Y^f = V^{f}_L \hat{Y}^f V^{f\,\dagger}_R,~\qquad f= u,d,e \,, 
\label{eq:rot}
\end{align}
where $\hat{Y}^f$ are diagonal Yukawa matrices, and $V_L^f$ and $V_R^f$ are unitary matrices corresponding to rotations of LH and RH fields, respectively. With these conventions, the CKM matrix reads $V_\text{CKM} = V_L^{u\,T} V_L^{d\,*}$. From Eq.~\eqref{eq:yukawas}, we see that the LH 1-2 rotations are $\ord{\varepsilon_\chi^2/\varepsilon_\phi^2}$ both in the up and down sector. Barring the effect of the $\ord{1}$ coefficients, the observed Cabibbo angle thus requires $\varepsilon_\chi/\varepsilon_\phi \approx 0.4 - 0.5$.
In general, imposing the experimental measurements of quark and charged lepton masses, as well as the hadronic and leptonic mixing angles, one can perform a numerical fit to the parameter set $\{ \lambda_{ij}^{u,d,e}, \varepsilon_{\phi}, \varepsilon_{\chi} \}$, as thoroughly discussed in~\cite{Falkowski:2015zwa,Linster:2018avp}. 
 
 Part of the calculation can be performed analytically, by perturbatively diagonalizing the Yukawa matrices ${Y}^f$~\cite{Falkowski:2015zwa}. 
 However, since the 2-3 rotation angles are large in both the RH down-type quark and the LH charged lepton sectors, we find it more straightforward to numerically diagonalize the Yukawa matrices for our analysis. In the following, we adopt the fit presented in~\cite{Calibbi:2020jvd} and numerically compute the singular value decomposition (SVD) of the matrices in Eq.~\eqref{eq:yukawas}, in order to obtain the field rotations $V_L^f$ and $V_R^f$. 
 For illustration, we show here the approximate form of such matrices:
\begin{align}
& V_L^u\sim V_R^u\sim
\begin{pmatrix}
 1& \frac{\varepsilon_{\chi}^{2}}{\varepsilon_{\phi}^{2}} &  \varepsilon_{\phi}\varepsilon_{\chi}^{2}
 \\
\frac{\varepsilon_{\chi}^{2}}{\varepsilon_{\phi}^{2}} & 1 & \varepsilon_{\phi} \\
 \frac{\varepsilon_{\chi}^{2}}{\varepsilon_{\phi}}& \varepsilon_{\phi} & 1
\end{pmatrix} \sim \begin{pmatrix}
1 & \lambda & \lambda^7 \\
\lambda & 1 & \lambda^2 \\
\lambda^3 & \lambda^2 & 1
\end{pmatrix}
, \nonumber \\
& V_L^d\sim V_R^e\sim
\begin{pmatrix}
1 & \frac{\varepsilon_{\chi}^{2}}{\varepsilon_{\phi}^{2}} & \frac{\varepsilon_{\chi}^{2}}{\varepsilon_{\phi}} \\
\frac{\varepsilon_{\chi}^{2}}{\varepsilon_{\phi}^{2}} & 1 & \varepsilon_{\phi} \\
\frac{\varepsilon_{\chi}^{2}}{\varepsilon_{\phi}} & \varepsilon_{\phi} & 1
\end{pmatrix}
\sim \begin{pmatrix}
1 & \lambda & \lambda^3 \\
\lambda & 1 & \lambda^2 \\
\lambda^3 & \lambda^2 & 1
\end{pmatrix},
\label{eq:rotations}
\\
\nonumber 
& V_R^d\sim V_L^e\sim
\begin{pmatrix}
1 & \frac{\varepsilon_\chi^2}{\varepsilon_\phi^2} & \varepsilon_\chi^2 \\
\frac{\varepsilon_\chi^2}{\varepsilon_\phi^2} & 1 & 1 \\
\frac{\varepsilon_\chi^2}{\varepsilon_\phi^2} & 1 & 1
\end{pmatrix}
\sim 
\begin{pmatrix}
1 & \lambda & \lambda^5 \\
\lambda & 1 & 1 \\
\lambda & 1 & 1
\end{pmatrix},
\end{align}
where according to the fit in~\cite{Calibbi:2020jvd}  
\begin{equation}
    \varepsilon_\chi = 0.008\,, \quad \varepsilon_\phi = 0.023\,,
    \label{eq:vevs-num}
\end{equation}
and $\lambda$ is a parameter of the order of the Cabibbo angle, $\lambda \approx \sin\theta_c \approx 0.2$. 

Finally, the neutrino mass matrix reads
\begin{equation}
m_{\nu}\approx\frac{v^{2}}{2\Lambda_L}
\begin{pmatrix}
\lambda_{11}^{\nu}\epsilon_{\phi}^{2}\epsilon_{\chi}^{4} & \lambda_{12}^{\nu}\epsilon_{\phi}^{2}\epsilon_{\chi}^{2} & \lambda_{13}^{\nu}\epsilon_{\phi}\epsilon_{\chi} \\
\lambda_{12}^{\nu}\epsilon_{\phi}^{2}\epsilon_{\chi}^{2} & \lambda_{22}^{\nu}\epsilon_{\phi}^{2} & \lambda_{23}^{\nu}\epsilon_{\phi}\epsilon_{\chi} \\
\lambda_{13}^{\nu}\epsilon_{\phi}\epsilon_{\chi} & \lambda_{23}^{\nu}\epsilon_{\phi}\epsilon_{\chi} & \lambda_{33}^{\nu}\epsilon_{\chi}^{2}
\end{pmatrix},
\end{equation}
where $v \simeq 246\,\mathrm{GeV}$ is the SM Higgs VEV.
As mentioned above, the structure of the neutrino mass matrix $m_\nu$ is obtained under the assumption that neutrinos are Majorana particles with masses arising from the Weinberg operator~\cite{Weinberg:1979sa} suppressed by the UV lepton number breaking scale $\Lambda_L$.

\subsection{$U(2)_F$ Nambu-Goldstone bosons}
\label{sec:goldstones}

The flavon VEVs displayed in Eq.~\eqref{eq:vevs} completely break the flavor symmetry implying the existence of four Nambu-Goldstone bosons associated with the four broken generators of $U(2)_F$. These become apparent when both scalar fields are decomposed in terms of their angular and radial components:
\begin{align}
 \phi(x) = \frac{e^{i\,\widetilde\pi_{i}{(x)}\,\sigma_{i}/v_{\phi}}}{\sqrt{2}}\begin{pmatrix}
{v_{\phi} + \varphi(x)} \\
0
\end{pmatrix}\,,
\quad
\chi(x) =\frac{1}{\sqrt2}\left(v_\chi+\rho\left(x\right)\right)e^{i \,\widetilde a(x)/v_\chi}\,, 
\label{eq:pngb1}
\end{align}
where $\sigma_{i}$ ($i = 1, 2, 3$) are the Pauli matrices, and $\widetilde \pi_{i}$, $\widetilde a$ represent the four Nambu-Goldstone modes. The radial modes $\varphi$ and $\rho$ naturally acquire masses of the order of the flavor symmetry breaking scale $v_{\phi} \sim v_{\chi}$, and their interactions with SM fermions are suppressed by factors $\sim m_f / v_{\phi,\chi}$. Hence, we will ignore their effects in the following.

Being associated with diagonal generators, the fields $\widetilde a$ and $\widetilde \pi_{3}$ mix, reflecting the fact that the $U(1)_F$ subgroup is broken by both $\left\langle \phi \right\rangle$ and $\left\langle \chi \right\rangle$, hence its associated Nambu-Goldstone mode is a linear combination of the phases of both flavons~\cite{Linster:2018avp}. Explicitly, we will work within the following physical basis:
\begin{equation}
\pi^{\prime}_1 = \widetilde \pi_1\,\quad \pi^{\prime}_2 = \widetilde \pi_2\,,\quad
\pi^\prime_{3}=\frac{v_\chi \widetilde \pi_{3}-v_\phi\widetilde a}{\sqrt{v_\chi^2+v_\phi^2}}\,, \quad 
a=\frac{v_{\chi}\widetilde a+v_{\phi}\widetilde \pi_{3}}{\sqrt{v_\chi^2+v_\phi^2}}\,,
\label{eq:pngb2}
\end{equation}
where we denoted as $\pi^\prime_i$ and $a$ the Nambu-Goldstone bosons associated, respectively, with $SU(2)_F$ and $U(1)_F$. This latter symmetry has a color anomaly, as is apparent from the quark charges shown in Table~\ref{tab:charges}, hence the field $a$ plays the role of the QCD axion~\cite{Linster:2018avp} and, in the following, we will refer to it as axiflavon~\cite{Calibbi:2016hwq}. On the other hand, $SU(2)_F$ is anomaly free and thus could be either local or global. In the following sections, we will discuss both options in turn, but let us first review the fundamental properties of the axiflavon.

\subsection{Axiflavon}
\label{sec:axiflavon}

As a consequence of the electromagnetic and color anomalies of $U(1)_F$, the axiflavon $a$ acquires interactions with photons and gluons, the latter of which enables the solution of the strong CP problem and, in general, makes $a$ behave like a QCD axion.
After performing the standard $a(x)$-dependent redefinitions of the fermion fields, such interactions\,---\,as well as the axiflavon couplings with fermions stemming from Eq.~\eqref{eq:yukawas}\,---\,can be written in the usual form:
\begin{align}
\label{eq:Laxiflavon}
{\cal L}_a  =  \frac{a}{f_a} \frac{\alpha_s}{8 \pi} G^a_{\mu \nu} \widetilde{G}^{a\,\mu \nu} + \frac{E}{N} \frac{a}{f_a} \frac{\alpha_{\rm em}}{8 \pi} F_{\mu \nu} \widetilde{F}^{\mu \nu}  +\frac{\partial_\mu a}{2 f_a} \,\overline{f}_i \gamma^\mu \left( C^V_{f_i f_j} + C^A_{f_i f_j} \gamma_5 \right) f_j \, ,
\end{align}
where, in our case, the anomaly coefficients are $E = 10$, $N = 9/2$, as follows from the charge assignments in Table~\ref{tab:charges}. In order to match the standard normalization of the interactions in Eq.~\eqref{eq:Laxiflavon}, the axion decay constant is defined as~\cite{Linster:2018avp} 
\begin{align}
f_a \equiv \frac {\sqrt{v_\chi^2 + v_\phi^2}}{\sqrt{2} N} = \frac {\sqrt{1 + \varepsilon_\chi^2/\varepsilon_\phi^2}}{\sqrt{2} N}\,v_\phi\, \simeq\, 0.17 \,v_\phi\,,
\label{eq:fa}
\end{align}
where we used the values of the expansion parameters shown in Eq.~\eqref{eq:vevs-num}.

The dimensionless coefficients that control the vector and axial axiflavon-fermion interactions in Eq.~\eqref{eq:Laxiflavon} can be expressed in terms of fermion charges and rotations as~\cite{Linster:2018avp}
\begin{align}
C^V_{f_i f_j} &= \frac{[f_{R\, a}] - [f_{L\, a}]}{2N} \delta_{ij} 
+ \frac{[f_{R \, 3}] - [f_{R \,a}]}{2N} \varepsilon^{f_R}_{ij} 
- \frac{[f_{L\, 3}] - [f_{L\, a}]}{2N} \varepsilon^{f_L}_{ij}\,, \\
C^A_{f_i f_j} &= \frac{[f_{R\, a}] + [f_{L\, a}]}{2N} \delta_{ij} 
+ \frac{[f_{R\, 3}] - [f_{R\, a}]}{2N} \varepsilon^{f_R}_{ij} 
+ \frac{[f_{L\, 3}] - [f_{L\, a}]}{2N} \varepsilon^{f_L}_{ij}\,,
\end{align}
where $[f_{L/R}]$ denotes the $U(1)_F$ charge of the corresponding LH or RH fermion and
\begin{equation}
\varepsilon^{f_L}_{ij} \equiv (V^f_{L})_{3i} (V^f_{L})^*_{3j}\,, \quad 
\varepsilon^{f_R}_{ij} \equiv (V^f_{R})_{3i} (V^f_{R})^*_{3j}\,.
\end{equation}

As we can see, the axiflavon has flavor-violating interactions, but the 1-2 transitions are protected by the $U(2)_F$ structure, so that they only arise from mixing with the third generation and are thus suppressed compared to the Abelian axiflavon discussed in~\cite{Calibbi:2016hwq}. Hence, in contrast to the latter case, astrophysical observables currently set more stringent constraints on $f_a$ than flavor processes for the $U(2)_F$ axiflavon, at the $f_a \gtrsim 10^8~\mathrm{GeV}$ level~\cite{Linster:2018avp}\,---\,see Section~\ref{sec:axion-pheno} for details.

The color anomaly also provides the axiflavon with a mass term that follows the standard QCD axion relation obtained from chiral perturbation theory and lattice QCD~\cite{GrillidiCortona:2015jxo}:
\begin{equation}
\label{eq:ma}
m_a \simeq 5.7\, \mu\text{eV} \,\left( \frac{10^{12}\, \mathrm{GeV}}{f_a} \right).
\end{equation}

In addition, again just like any QCD axion, the energy density stored in the oscillations of the axiflavon field can account for the dark matter (DM) abundance observed in the present universe~\cite{Preskill:1982cy,Abbott:1982af,Dine:1982ah}. If the $U(1)_F$ breaking occurs before inflation, the resulting relic density reads
\begin{align}
\Omega_{a}h^{2}\simeq 0.12
\left(\frac{f_a}{9.5\times 10^{11}~\mathrm{GeV}}\right)^{7/6}\theta^{2}
\simeq 0.12
\left(\frac{v_\phi}{5.6\times 10^{12}~\mathrm{GeV}}\right)^{7/6}\theta^{2}
\,,
\end{align}
where $\theta$ denotes the initial misalignment angle. This implies that the misalignment mechanism can account for the observed DM relic density\,---\,$\Omega_\text{DM} h^2 \simeq 0.12$~\cite{Planck:2018vyg}\,---\,for naturally large values of the misalignment angle, $\theta  \approx 0.1-1$, if the flavor symmetry is broken at a scale \mbox{$v_\phi  \approx 10^{12}-10^{14}~\mathrm{GeV}$}.

\section{$SU(2)_F$ gauge bosons}
\label{sec:local}

As mentioned above, one can consider the possibility that $U(1)_F$ is global while $SU(2)_F$ is a gauge symmetry, as the latter (unlike $U(1)_F$) is anomaly free.\footnote{Witten anomaly is also absent if the number of $SU(2)$ doublets is even~\cite{Witten:1982fp}. Hence, in our case, the introduction of the RH neutrino doublet $N_a$ is sufficient to consistently gauge $SU(2)_F$.} In this scenario, the three Nambu-Goldstone bosons $\widetilde \pi_i$ introduced in Section~\ref{sec:goldstones} become the longitudinal components of the $SU(2)_F$ gauge boson triplet that we label $W^\prime$. The relevant part of the Lagrangian for the local case is given by:
\begin{equation}
\mathcal{L} \supset \left| D_{\mu} \phi \right|^2  + \left| \partial_{\mu} \chi \right|^2 + \frac{1}{4} W_{\mu\nu}^{\prime}\cdot W^{\prime\,\mu\nu} - V(\phi, \chi)\,,
\end{equation}
where $V(\phi, \chi)$ is the scalar potential involving the flavons,\footnote{For simplicity, we assume that scalar potential terms inducing mixing with the Higgs doublet are negligible.} and the covariant derivative reads 
\begin{equation}
D_{\mu} = \partial_{\mu} + i g_F \, \frac{\sigma \cdot W_{\mu}^{\prime}}{2}\,,
\end{equation}
such that the mass term of the gauge bosons $W_{i\,\mu}^{\prime}$ ($i=1,2,3$) results
\begin{equation}
-\mathcal{L}_{W^\prime} \supset \frac{1}{8} g_F^2  v_{\phi}^2 \left( W_{\mu}^{\prime} \cdot W^{\prime\,\mu} \right)=\frac{1}{2}m^2_{W^{\prime}}\left( W_{\mu}^{\prime} \cdot W^{\prime\mu} \right) \quad \Rightarrow \quad m_{W^\prime} = \frac{1}{2} g_F v_\phi\,,    
\label{eq:mWp}
\end{equation}
where $g_F$ is the $SU(2)_F$ gauge coupling.
\revision{Hence, we see that $W^\prime$ comprises three degenerate electromagnetically neutral states.}

The SM fermions belonging to the $SU(2)_F$ doublets in Table~\ref{tab:charges} directly couple to the $W^{\prime}$ bosons through the covariant derivative in the respective kinetic terms. The third-generation quarks and leptons have no coupling with $W^{\prime}$ in the interaction basis, as they are singlets. However, third-generation mass eigenstates still acquire such couplings through the rotations shown in Eq.~\eqref{eq:rotations}.
In the interaction basis, the couplings between $SU(2)_F$ gauge bosons and SM fermions can be written as:
\begin{equation}
\mathcal{L}_{W^\prime ff} = \frac{g_F}{2} 
W^{\prime}_{i\,\mu} \left( 
\overline{f_{L}}_\alpha C_{\alpha\beta}^{i} \gamma^{\mu} f_{L\,\beta} 
+ 
\overline{f_{R}}_\alpha C_{\alpha\beta}^{i} \gamma^{\mu} f_{R\,\beta} 
\right),
\end{equation}
where $\alpha,\beta = 1,2,3$ are flavor indices and the matrices $C^{i}$ read
\begin{equation}
C^{1} = \begin{pmatrix}
0 & 1 & 0 \\
1 & 0 & 0 \\
0 & 0 & 0
\end{pmatrix},\quad
C^{2} = \begin{pmatrix}
0 & -i & 0 \\
i & 0 & 0 \\
0 & 0 & 0
\end{pmatrix},\quad
C^{3} = \begin{pmatrix}
1 & 0 & 0 \\
0 & -1 & 0 \\
0 & 0 & 0
\end{pmatrix}.
\end{equation}
These matrices are in the form of expanded Pauli matrices, reflecting the $SU(2)_F$ symmetry structure. Note that $C^1$ and $C^2$ have non-vanishing 12 and 21 entries, implying maximal (unsuppressed) flavor violation between first and second generation fermions. After performing fermion field rotations defined in Eq.~\eqref{eq:rot}, we obtain the coupling coefficients in the mass basis, $C^{f\,i}_{L/R} \equiv V_{L/R}^{f\,\dagger} C^i V_{L/R}^f$. For example, the couplings of LH down quarks result:
\begin{align}
&C^{d\,1}_{L}  = V_{L}^{d\,\dagger} C^1 V^d_{L} \sim 
\begin{pmatrix}
\lambda & 1 & \lambda^2\\
1 & \lambda & \lambda^3 \\
\lambda^2 & \lambda^3 & \lambda^5
\end{pmatrix},\quad
C^{d\,2}_{L}  = V_{L}^{d\,\dagger} C^2 V^d_{L} \sim 
\begin{pmatrix}
0 & 1 & \lambda^2\\
1 & 0 & \lambda^3 \\
\lambda^2 & \lambda^3 & 0
\end{pmatrix}, \nonumber\\
&C^{d\,3}_{L}  = V_{L}^{d\,\dagger} C^3 V^d_{L} \sim 
\begin{pmatrix}
1 & \lambda & \lambda^3\\
\lambda & 1 & \lambda^2 \\
\lambda^3 & \lambda^2 & \lambda^4
\end{pmatrix}.
\label{eq:Cdmatr}
\end{align}
We can then write the $W^\prime$ interactions with vector and axial currents:
\begin{equation}
\mathcal{L}_{W^\prime ff}  = \frac{g_{F}}{2}W_{i\,\mu}^{\prime} 
\left[\overline{f}_{\alpha} \gamma^{\mu} 
\left(C_{V\,\alpha\beta}^{f\,i} + C_{A\,\alpha\beta}^{f\,i} \gamma_{5}\right) f_{\beta}\right], \quad 
C_{V,A}^{f\,i} = \frac{C_{R}^{f\,i} \pm C_{L}^{f\,i}}{2}\,.
\label{eq:LWff}
\end{equation}

As we have seen, flavor-violating $W^{\prime}$ couplings arise naturally from the structure of the $SU(2)_F$ symmetry and exhibit peculiar hierarchical patterns due to the rotation matrices of the fermion fields. As a result, the $W^{\prime}$ gauge bosons mediate flavor-changing neutral current (FCNC) processes and lepton-flavor-violating (LFV) decays, such as $K^0 - \overline{K^0}$ oscillations, $\mu \to eee$ etc. If the $W^{\prime}$ triplet is sufficiently light, exotic meson decays, including \mbox{$K \to \pi W^{\prime}$} and $B \to K W^{\prime}$, and leptonic decays, such as $\mu \to e W^{\prime}$, must be included in the analysis, as they will pose stringent constraints on the model. Expressions for the above processes and the resulting limits on the flavor breaking scale are presented in the phenomenological discussion in Section~\ref{sec:pheno}.

In order to assess whether the $W^\prime$ gauge bosons give rise to a visible signal in a typical particle detector, we need to estimate their lifetime. 
In terms of the couplings defined in Eq.~(\ref{eq:LWff}), when kinematically allowed, the widths for the $W^{\prime}$ decays into fermions read:
\begin{align}
\label{eq:Wpwidth}
& \Gamma(W^{\prime}_i \to f_{\alpha}\overline{f}_{\beta})
=\frac{N_{c}^{f}g_{F}^{2} \, m_{W^{\prime}}}{12\pi}\sqrt{\left(1-\frac{(m_{f_{\alpha}}+m_{f_{\beta}})^{2}}{m_{W^{\prime}}^{2}}\right)\left(1-\frac{(m_{f_{\alpha}}-m_{f_{\beta}})^{2}}{m_{W^{\prime}}^{2}}\right)} \, \times  \nonumber \\
 & \left[\left(1-\frac{m_{f_{\alpha}}^{2}+m_{f_{\beta}}^{2}}{2m_{W^{\prime}}^{2}}\right)\left(|C_{V\,\alpha\beta}^{f\,i}|^{2}+|C_{A\,\alpha\beta}^{f\,i}|^{2}\right)\!+3\frac{m_{f_{\alpha}}m_{f_{\beta}}}{m_{W^{\prime}}^{2}}\left(|C_{V\,\alpha\beta}^{f\,i}|^{2}-|C_{A\,\alpha\beta}^{f\,i}|^{2}\right)\right]\,, 
\end{align}
where $N^f_c$ is the color factor of the fermion $f$. 
For heavy $W^\prime$, the total decay width is then 
\begin{equation}
\Gamma_{W^{\prime}_i} = \sum_{\alpha, \beta} \left[ \Gamma(W^{\prime}_i \to u_{\alpha} \overline{u}_{\beta}) + \Gamma(W^{\prime}_i \to d_{\alpha} \overline{d}_{\beta}) + \Gamma(W^{\prime}_i \to \ell_{\alpha} \overline{\ell}_{\beta}) + \Gamma(W^{\prime}_i \to \nu_{\alpha} \overline{\nu}_{\beta}) \right]\,.
\end{equation}
It is understood that contributions below the kinematic threshold have to be eliminated.
This expression is valid only for $m_{W^{\prime}}\gg \Lambda_\text{QCD}$. Nevertheless, for simplicity, we are going to employ it also to estimate the lifetime of a lighter $W^{\prime}$, hence neglecting hadronization effects.\footnote{In addition, decays of massive vectors into three photons or gluons are induced by fermion loops at highly suppressed rates\,---\,see e.g.~\cite{Redondo:2008ec}\,---\,such that they can be safely neglected in the kinematic regime we consider in the following, where at least $W_i^{\prime}\to e^+e^-$ is always kinematically allowed.} Still, for light quarks, no contribution from decays into up and down (strange) quarks is included below the pion (kaon) threshold.
Notice that the QCD effects we are neglecting become particularly relevant for $m_{W^\prime} \lesssim \Lambda_\text{QCD} \approx 1\,\mathrm{GeV}$. On the other hand, as we will see, light gauge bosons\,---\,especially below the threshold of $W^\prime \to \mu e$, that is, for $m_{W^\prime} \lesssim m_\mu $\,---\,are always long-lived enough to escape detection in a typical particle physics experiment detector, hence the precise value of their lifetime does not affect our analysis. In other words, our simplifying assumption is a source of inaccuracy only within the limited regime $0.1\,\mathrm{GeV}\lesssim m_{W^\prime} \lesssim 1\,\mathrm{GeV}$.

\section{$SU(2)_F$ pseudo Nambu-Goldstone bosons}
\label{sec:global}

We now introduce the second case of a global $SU(2)_F$ featuring, besides the axiflavon, the three additional Nambu-Goldstone bosons $\pi^\prime_i$ defined in Eqs.~\eqref{eq:pngb1} and \eqref{eq:pngb2}. In the following, we assume them to be massive PNGBs due to small explicit $SU(2)_F$ breaking. Specifically,
we introduce an explicit soft symmetry-breaking term in the scalar potential:
\begin{equation}
\delta V = m_{\pi^\prime}^2 (\phi^\dagger \sigma_3 \phi)\,,
\end{equation}
where $\sigma_3$ denotes the third Pauli matrix. Substituting Eq.~\eqref{eq:pngb1} into this expression, mass terms for the $SU(2)_F$ PNGBs arise such that $m_{\pi^{\prime}_1} = m_{\pi^{\prime}_2} = m_{\pi^\prime}$, $m_{\pi^{\prime}_3} = m_{\pi^\prime}/\sin\theta$ where we have taken into account the mixing ($\sin\theta \equiv v_\chi/\sqrt{v_\phi^2+v_\chi^2}$) with $U(1)_F$ PNGB as in Eq.~\eqref{eq:pngb2}. We can also assume this breaking term to arise from Planck-suppressed operators of the form $(\phi^\dagger \sigma_3 \phi)(\phi^\dagger \phi)^n$ with $n>1$, such that $m_{\pi^\prime}^2 \sim (v_\phi)^{2n}/M_\text{Pl}^{2(n-1)}$, where $M_\text{Pl}\simeq 1.22\times 10^{19}~\mathrm{GeV}$ denotes the Planck scale. This assumption is consistent with the commonly accepted notion that quantum gravity effects break any global symmetry\,---\,cf.~\cite{Reece:2023czb} for a recent review. In either case, we treat $m_{\pi^\prime}$ as a free parameter throughout the rest of the paper.

By substituting the full form of the flavon $\phi$, written as in Eq.~\eqref{eq:pngb1}, into the effective Yukawa Lagrangian, Eqs.~\eqref{eq:Lu}-\eqref{eq:Le}, one can obtain the interactions between the PNGBs and the SM fermions. It is then convenient to change the field basis by performing a field-dependent transformation of the fermion flavor doublets, for instance:
\begin{equation}
Q_a \to e^{i (\widetilde\pi_1\sigma_1+\widetilde\pi_2\sigma_2+\widetilde\pi^{\prime}_3\sigma_3)_{ab} / v_{\phi}}e^{-i\widetilde a/v_{\chi}} Q_b \quad (a,b=1,2),
\end{equation}
(with $\widetilde\pi_{3}{(x)}\,/v_{\phi}=\widetilde\pi_{3}^{\prime}{(x)}\,/v_{\phi}+\widetilde a{(x)}\,/v_{\chi}$) that removes the dependence on the PNGBs from the Yukawa sector. 
After further rotating the fermion fields to the mass basis, the fermion kinetic terms lead to the following PNGB-fermion couplings: 
%
%
\begin{equation}
\mathcal{L}_{\pi^{\prime} ff} = \frac{\partial_{\mu} \pi^{\prime}_i}{v_{\phi}} \left[ \overline{f}_{\alpha} \gamma^{\mu} \left(\hat C_{V\,\alpha\beta}^{f\,i} + \hat C_{A\,\alpha\beta}^{f\,i} \gamma_5 \right) f_{\beta} \right]\,,
\label{eq:Lpiff}
\end{equation}
where the coupling matrices $\hat C^{f\,1,2}_{V/A} = C^{f\,1,2}_{V/A}$, that is, they are the same as those appearing in Eq.~(\ref{eq:LWff}), while the couplings of $\pi^\prime_3$ are modified by the mixing in Eq.~\eqref{eq:pngb2} such that:
\begin{align}
\hat C^{f\,3}_{L/R} = V_{L/R}^{f\,\dagger} \begin{pmatrix}
2/\sin\theta - \sin\theta & 0 & 0 \\
0 & -\sin\theta & 0 \\
0 & 0 & 0
\end{pmatrix}
 V_{L/R}^f\,, \quad \hat C_{V,A}^{f\,3} = \frac{\hat C_{R}^{f\,3} \pm \hat C_{L}^{f\,3}}{2}\,,
 \label{eq:picouplg}
\end{align}
where $\sin\theta \equiv {v_\chi}/ {\sqrt{v_\phi^2+v_\chi^2}} \simeq 0.32$.

In terms of the above couplings, the widths of the kinematically allowed $\pi^\prime_i$ decays read
\begin{equation}
\label{eq:pipwidth}
\Gamma(\pi^{\prime}_i \to f_{\alpha} \bar f_{\beta}) = \frac{N_{c}^{f} m_{\pi^{\prime}}}{8\pi} \left[ \left( \frac{m_{f_{\alpha}} - m_{f_{\beta}}}{v_{\phi}} \right)^2 |\hat C_{V\,\alpha\beta}^{f\,i}|^2 z_+ + \left( \frac{m_{f_{\alpha}} + m_{f_{\beta}}}{v_{\phi}} \right)^2 |\hat C_{A\,\alpha\beta}^{f\,i}|^2 z_- \right] \sqrt{z_+ z_-}\,,
\end{equation}
where $N_{c}^{f}$ is the color factor and $z_{\pm} = 1 - {(m_{f_{\alpha}} \pm m_{f_{\beta}})^2}/{m_{\pi^{\prime}}^2}$.
For heavy PNGBs, the total decay width is thus given by:
\begin{equation}
\Gamma_{\pi^{\prime}_i} = \sum_{\alpha, \beta} \left[ \Gamma(\pi^{\prime}_i \to u_{\alpha} \overline{u}_{\beta}) + \Gamma(\pi^{\prime}_i \to d_{\alpha} \overline{d}_{\beta}) + \Gamma(\pi^{\prime}_i \to \ell_{\alpha} \overline{\ell}_{\beta}) + \Gamma(\pi^{\prime}_i \to \nu_{\alpha} \overline{\nu}_{\beta}) \right],
\end{equation}
where we neglected loop-induced couplings to SM gauge bosons, as we are always working in a regime where some of the above decays into fermions are kinematically open. Similarly to the $W^\prime$ case discussed in the previous section, we use this perturbative expression to estimate the order of magnitude of the $\pi^\prime$ lifetime also for values of $m_{\pi^\prime}$ for which hadronization effects should be taken into account, expecting this simplifying assumption to be source of inaccuracy only within the limited regime $0.1\,\mathrm{GeV}\lesssim m_{\pi^\prime} \lesssim 1\,\mathrm{GeV}$.


\section{Phenomenological implications of the new flavored bosons}
\label{sec:pheno}

We now discuss the most relevant constraints from low-energy FCNC and LFV processes on the considered $U(2)_F$ flavor model. 
To summarize, in the global $SU(2)_F$ symmetry case, the theory features four PNGBs,  $a$ (the axiflavon) and $\pi^{\prime}_i$ ($i = 1, 2, 3$), whereas in the local symmetry case, the new bosons are the gauge bosons $W^{\prime}_i$ ($i = 1, 2, 3$) and again the axiflavon~$a$. When discussing general effects, we refer to these fields collectively as $X$.

In both scenarios, flavor-violating couplings of the $U(2)_F$ bosons with SM fermions are naturally induced by the structure of the symmetry. In particular, $SU(2)_F \subset U(2)_F$ leads to unsuppressed flavor violation involving the first two fermion generations. In addition, flavor-violating processes have a mild dependence on the $\ord{1}$ coefficients of the model and, in the case of those mediated or involving the axiflavon, the $U(1)_F$ charges of the SM fermions.\footnote{\revision{In order to evaluate a given process, the couplings appearing in the Lagrangians in the previous sections should be evolved from the symmetry-breaking scale down to the energy scale characteristic of the process of interest. The renormalization group equations for the couplings of axions and ALPs are, for instance, given and their numerical impact is studied in~\cite{Chala:2020wvs,Bauer:2020jbp,Bisht:2024hbs}. However, for the flavor-violating interactions that we are interested in, running effects are suppressed by small Yukawa couplings and/or CKM angles. In agreement with the numerical estimates in~\cite{Bauer:2020jbp}, we find that such corrections are thus always below the 10\% level, hence we neglect them throughout our analysis.}}

\subsection{Decays into light $SU(2)_F$ bosons}
\label{sec:light-fcnc}

If the mass of the boson $X$ is light enough, it can be directly produced on-shell in flavor-violating meson or lepton decays, such as $K \to \pi X$, $B \to KX$, $\mu \to eX$. This situation is rather natural in the global $SU(2)_F$ scenario, $X = \pi^\prime_i$, because of the PNGB nature of $\pi^\prime_i$, while, in the local $SU(2)_F$ case, $X = W^\prime_i$, this requires $g_F \ll 1$, as can be seen from Eq.~\eqref{eq:mWp}.

If kinematically allowed, the above exotic two-body  decays set some of the strongest constraints on the $U(2)_F$ breaking scale, as we will see below, and can provide spectacular experimental signatures of the model under study.

\paragraph{Gauge bosons.}
In the local case, adapting the expressions in~\cite{Smolkovic:2019jow,Eguren:2024oov}, we obtain for the branching ratios of the most relevant meson decays into light gauge bosons:
\begin{align}
\label{eq:KpiWp}
\text{BR}(K^{+} \to \pi^{+} W^{\prime}_i)~ &=~ \frac{g_{F}^2}{64 \pi\, m_{W^{\prime}}^2}  \frac{m_K^3}{\Gamma_K} \left[ \lambda \left( 1, \frac{m_{\pi}^2}{m_K^2}, \frac{m_{W^{\prime}}^2}{m_K^2} \right) \right]^{3/2} \left[ f^K_+ \left( m_{W^{\prime}}^2 \right) \right]^2 \left| C_{V\,sd}^{d\,i} \right|^2\,,
\\
\label{eq:BKWp}
\text{BR}(B^{+} \to K^{+} W^{\prime}_i~) &=~ \frac{g_{F}^2}{64 \pi \,m_{W^{\prime}}^2}  \frac{m_B^3}{\Gamma_B} \left[ \lambda \left( 1, \frac{m_K^2}{m_B^2}, \frac{m_{W^{\prime}}^2}{m_B^2} \right) \right]^{3/2} \left[ f^B_+ \left( m_{W^{\prime}}^2 \right) \right]^2 \left|C_{V\,bs}^{d\,i}  \right|^2\,,
\end{align}
where $\lambda(x,y,z) \equiv x^2 + y^2 + z^2 - 2(xy + yz + xz)$, the meson widths are $\Gamma_K \simeq 5.3 \times 10^{-17} \, \mathrm{GeV}$ and $\Gamma_B \simeq 4.0 \times 10^{-13} \, \mathrm{GeV}$ \cite{ParticleDataGroup:2022pth}, and the relevant hadronic form factors $f^{K,B}_+$ can be found in~\cite{FlavourLatticeAveragingGroup:2019iem,Bailey:2015dka}.
The couplings appearing in these expressions are those to vector quark currents given in Eq.~\eqref{eq:LWff}.

Similarly, we can derive the branching ratios of LFV decays such as $\mu \to e W^\prime_i$ from the expressions in \cite{Smolkovic:2019jow,Heeck:2016xkh,Ibarra:2021xyk}, obtaining:
\begin{equation}
\label{eq:ellellWp}
\text{BR}(\ell_\alpha \to \ell_\beta \,W^{\prime}_i) ~=~ \frac{g_F^2}{64 \pi\,m_{W^{\prime}}^2}  \frac{m_{\ell_\alpha}^3}{\Gamma_{\ell_\alpha}} \left( \left| C_{V\,\alpha\beta}^{e\,i} \right|^2 + \left| C_{A\,\alpha\beta}^{e\,i} \right|^2 \right) \left( 1 + 2 \frac{m_{W^{\prime}}^2}{m_{\ell_\alpha}^2} \right) \left( 1 - \frac{m_{W^{\prime}}^2}{m_{\ell_\alpha}^2} \right)^2\,,
\end{equation}
where the mass of the lighter lepton was neglected because $m_{\ell_\beta}\ll m_{\ell_\alpha}$, and the total lepton widths are
$\Gamma_{\mu} \simeq 3.0\times 10^{-19} \, \mathrm{GeV}$, 
$\Gamma_{\tau} \simeq 2.3 \times 10^{-12} \, \mathrm{GeV}$~\cite{ParticleDataGroup:2022pth}.

Given the standard relation among the $W^\prime$ mass, the VEV, and the gauge coupling in Eq.~\eqref{eq:mWp}, $g_F$ drops from the prefactor in the above expressions such that the latter only depends on the $U(2)_F$ breaking scale, resulting in $1/(16\pi\, v_\phi^2)$.

\paragraph{PNGBs.}
In the global case, we can use results from the literature on flavour-violating axions and axion-like particles (ALPs), e.g.~\cite{MartinCamalich:2020dfe,Bauer:2021mvw,Altmannshofer:2023hkn}, obtaining the following expressions for the branching ratios of the relevant meson decays:
\begin{align}
 \text{BR}(K^{+} \to \pi^{+} \pi^{\prime}_i) ~& = ~ \frac{m_K^3}{16\pi \,v_{\phi}^2\, \Gamma_K }\left(1-\frac{m_{\pi}^2}{m_K^2}\right)^2\left[ \lambda \left( 1, \frac{m_{\pi}^2}{m_K^2}, \frac{m_{\pi^{\prime}_i}^2}{m_K^2} \right) \right]^{1/2}\left[ f^K_0 \left( m_{\pi^{\prime}_i}^2 \right) \right]^2\left| \hat{C}_{V\,sd}^{d\,i} \right|^2\,, 
\\
 \text{BR}(B^{+} \to K^{+} \pi^{\prime}_i)~& =~ \frac{m_B^3}{16\pi \,v_{\phi}^2\,\Gamma_B}\left(1-\frac{m_K^2}{m_B^2}\right)^2\left[ \lambda \left( 1, \frac{m_{K}^2}{m_B^2}, \frac{m_{\pi^{\prime}_i}^2}{m_B^2} \right) \right]^{1/2}\left[ f^B_0 \left( m_{\pi^{\prime}_i}^2 \right) \right]^2\left| \hat{C}_{V\,bs}^{d\,i} \right|^2\,,
\end{align}
where the form factors $f^{K,B}_0$ are also provided by~\cite{FlavourLatticeAveragingGroup:2019iem,Bailey:2015dka}, and the couplings $\hat{C}_{V\,\alpha\beta}^{d\,i}$ are given in Eqs.~\eqref{eq:Lpiff} and \eqref{eq:picouplg}.

In the leptonic sector, we can adapt the expressions for LFV decays into ALPs~\cite{Cornella:2019uxs,Calibbi:2020jvd,Bauer:2021mvw}:
\begin{equation}
\text{BR}(\ell_\alpha\to\ell_\beta\,\pi^\prime_i) ~=~ \frac{1}{16\pi\,v_{\phi}^2}\frac{m_{\ell_\alpha}^3}{\Gamma_{\ell_\alpha}}\left(|\hat{C}_{V\,\alpha\beta}^{e\,i}|^2+|\hat{C}_{A\,\alpha\beta}^{e\,i}|^2\right)\left(1-\frac{m_{\pi^\prime_i}^2}{m_{\ell_\alpha}^2}\right)^2\,.
\end{equation}

Notice that the prefactors in the above expressions are the same as those appearing in Eqs.~\eqref{eq:KpiWp}-\eqref{eq:ellellWp}, so that the decay rates into PNGBs and gauge bosons are identical in the limits \mbox{$m_{W^\prime} \ll m_P$}, $m_{\pi^\prime} \ll m_P$ (with $m_P$ denoting the mass of the parent particle), consistent with the Goldstone equivalence theorem.

\paragraph{Light bosons decays and decay length.}
The lifetime of the light particle $X$ plays a crucial role in determining its experimental signatures, and thus which experiments are sensitive to the above production modes and what the resulting constraints are. The probability that $X$ decays within the detector volume is
\begin{equation}
1 - \exp\left(-\frac{L}{l_X}\right)\,,
\end{equation}
where $L$ represents the detector size (e.g.~$L\approx 1\,\mathrm{m}$ for typical LFV experiments~\cite{Bernstein:2013hba,Calibbi:2017uvl}, $L\approx 100\,\mathrm{m}$ for kaon experiments such as NA62~\cite{NA62:2017rwk}, etc.), and $l_X$ is the mean decay length of $X$ in the laboratory frame that, in natural units, is given by
\begin{equation}
l_X = \frac{p_X}{m_X \Gamma_X}\,, \qquad 
p_X = \frac{\sqrt{\lambda\left(m_{P}^2, m_{P^\prime}^2, m_X^2\right)}}{2 m_{P}}\,,
\end{equation}
where $m_{P^{(\prime)}}$ are the masses of the SM particles involved in the production process $P\to P^\prime\,X$.

In the following, we focus on the leptonic decays $X \to \ell \ell^{(\prime)}$. $\ell, \ell^{\prime} = e,\mu$, which are the main allowed channels in the regime $m_X \lesssim 1\,\mathrm{GeV}$ and the cleanest otherwise.
For a given value of its mass $m_X$, the average decay length of $X$ depends on its couplings to fermions. If the latter are sufficiently large, $X$ will decay promptly within the detector, producing visible dilepton signatures. In contrast, for small couplings, $X$ may be long-lived and escape the detector before decaying, resulting in a missing energy signal.\footnote{\revision{The viability and potential impact of the intermediate regime\,---\,corresponding to displaced-vertex decays\,---\,strongly depend on the specific process and detector. See Appendix~\ref{app:explim} for more detailed comments.}}

Consequently, different types of experimental constraints arise from semi-invisible two-body processes such as 
\begin{equation}
K \to \pi \,X_\text{inv}\,,\quad B \to K(\pi) \,X_\text{inv}\,,\quad \mu \to e\,X_\text{inv}\,,\quad \tau \to \ell \,X_\text{inv}\,,
\label{eq:invisible}
\end{equation}
or fully visible final states 
\begin{equation}
K \to \pi X (\to\ell \ell^{(\prime)})\,,\quad B \to K(\pi) X (\to\ell \ell^{(\prime)})\,,\quad \mu \to e X(\to ee)\,,\quad \tau \to \ell X (\to\ell \ell^{(\prime)})\,,
\label{eq:visible}
\end{equation}
where $ \ell \ell^{(\prime)} = e^+e^-,\, \mu^+ \mu^-,\, \mu^\pm e^\mp$, depending on the available phase space.

Denoting mesons or leptons by $P$ and $P^\prime$ (with $m_P > m_{P^\prime}$), the experimental constraints on the above processes can be obtained from the following expressions:
\begin{align}
&\exp\left(-\frac{L}{l_X}\right) \, \mathrm{BR}(P \to P^\prime\, X) 
~<~ \mathrm{BR}(P \to P^\prime + \text{inv})^{\text{exp}}\,, \\
&\left[1 - \exp\left(-\frac{\revision{L_\text{prompt}}}{l_X}\right)\right]
\, \mathrm{BR}(P \to P^\prime \,X) \, \mathrm{BR}(X \to \ell\ell^{(\prime)}) 
~<~ \mathrm{BR}(P \to P^\prime \, \ell\ell^{(\prime)})^{\text{exp}}\,,
\end{align}
\revision{where $L_\text{prompt}$ represents the decay length under which the decay appears prompt, which depends on the spatial resolution of a given detector. For simplicity, we take $L_\text{prompt} = 1\,\mathrm{mm}$ throughout our analysis with the exception of kaon processes, for which it is sufficient that $X$ decays within the detector to give rise to a visible signature.}
The above formulae enable us to confront our model's predictions with existing experimental bounds on processes with both visible and invisible final states\,---\,bounds that are respectively denoted as $\mathrm{BR}(P \to P^\prime + \text{inv})^{\text{exp}}$ and $\mathrm{BR}(P \to P^\prime \, \ell\ell^{(\prime)})^{\text{exp}}$. The experimental limits we use in our phenomenological study are presented in Appendix~\ref{app:explim}.

\subsection{Flavor violation mediated by heavy $SU(2)_F$ bosons}
\label{sec:heavy-fcnc}

$U(2)_F$ bosons that are heavy enough for the above decays to be kinematically forbidden can still mediate processes like those in Eq.~\eqref{eq:visible} off-shell. In addition, their flavor-violating interactions contribute to meson oscillations at the tree level with $K^0 - \overline{K^0}$ mixing providing the strongest quark sector constraint in this regime.

\paragraph{Gauge bosons.}
Let us start considering quark FCNC processes in the local case, again focusing on semi-leptonic meson decays, $M_1 \to M_2 \,\ell_\alpha \ell_\beta$ from the underlying quark transition \mbox{$q_a \to q_b$} (with $q=u$ or $d$).
Heavy $W^{\prime}$ bosons with mass $m_{W^{\prime}} \gg m_{M_1}$ can be integrated out resulting in effective dimension-six interactions with coefficients $\sim g_F^2 / 4m_{W^{\prime}}^2 = 1/v_\phi^2$. 
Following~\cite{Smolkovic:2019jow}, we write the resulting differential decay width in the approximation $m_{M_2},\, m_{\ell_\alpha},\, m_{\ell_\beta} \to 0$:
\begin{align}
\frac{\mathrm{d^2}\Gamma(M_1 \to M_2\, \ell_\alpha \bar\ell_\beta)}{\mathrm{d}m_{M_1\ell_\alpha}^2\, \mathrm{d}m_{\ell_\alpha\ell_\beta}^2}
\simeq &~
\frac{g_F^4  \left( \left|\sum_i C_{V\,ab}^{q\,i}C_{V\,\alpha\beta}^{e\,i}\right|^2 + \left|\sum_i C_{V\,ab}^{q\,i}C_{A\,\alpha\beta}^{e\,i}\right|^2 \right) }{512 \pi^3 m_{W^{\prime}}^4 m_{M_1} }\times\nonumber\\ 
& ~\left[f^{M_1}_+(m_{\ell_\alpha\ell_\beta}^2)\right]^2 m_{M_1\ell_\alpha}^2 \left(1 - \frac{m_{M_1\ell_\alpha}^2}{m_{M_1}^2} - \frac{m_{\ell_\alpha\ell_\beta}^2}{m_{M_1}^2} \right),
\end{align}
where $m_{M_1\ell_\alpha}^2 \equiv (p_{M_1} + p_{\ell_\alpha})^2$ and $m_{\ell_\alpha\ell_\beta}^2 \equiv (p_{\ell_\alpha} + p_{\ell_\beta})^2$ is the invariant mass squared of the dilepton pair, and $f^{M_1}_+$ denotes the hadronic form factors also appearing in Eq.~\eqref{eq:KpiWp} and \eqref{eq:BKWp}. Expressions with full dependence on the final state masses and valid also for a light mediator can be found in~\cite{Smolkovic:2019jow}.

Similarly, integrating out the $W^\prime$ triplet, we can obtain the Wilson coefficients (WCs) for the dimension-six $\Delta S = 2$ operators inducing neutral kaon oscillations:
\begin{equation}
(\overline{s}_X \gamma^\mu d_X)(\overline{s}_Y \gamma_\mu d_Y): \quad \sum_i\frac{g_F^2}{4 m_{W^{\prime}}^2}  C_{X\,sd}^{d\,i} C_{Y\,sd}^{d\,i} \,,
\end{equation}
where $X,Y = L,R$ and the LH and RH couplings can be calculated as illustrated in Section~\ref{sec:local}. In the numerical analysis, we also consider the light mediator case adapting the results of~\cite{Smolkovic:2019jow}, which, however, leads to negligible constraints compared to the processes discussed above in Section~\ref{sec:light-fcnc}.
Again, the WCs of the above operators are suppressed by a factor $1/v_{\phi}^2$, hence constraints from $K^0-\overline{K^0}$ mixing set a lower limit on the $SU(2)_F$ symmetry breaking scale. By comparing the above prediction with limits on the real part of the WCs from the measured kaon mass splitting $\Delta m_K$ that are reported in the literature~\cite{Silvestrini:2018dos}, we find that the most stringent constraint arises from the $W^\prime$ contribution to $(\overline{s}_L \gamma^\mu d_L)(\overline{s}_R \gamma_\mu d_R)$ and numerically results
\begin{equation}
\Delta m_K:\quad v_\phi \gtrsim 5.4 \times 10^6 \, \mathrm{GeV} \quad (\text{local case})\,.
\label{eq:DmKlocal}
\end{equation}
As we will see below, this limit is still orders of magnitude weaker than the limits one obtains if SM particles can decay into a $W^\prime$, but it represents the irreducible flavor bound on the symmetry breaking scale in the local case.

In the lepton sector, three-body LFV decays, such as $\mu \to eee$ and $\tau \to \mu\mu\mu$, are also induced at the tree level and are still suppressed only by the scale $v_\phi$, hence resulting phenomenologically more important than the corresponding radiative (loop induced) processes $\mu\to e\gamma$ or $\tau\to \mu\gamma$. In the heavy ${W^\prime}$ limit, the branching ratios of these processes can again be obtained from effective field theory expressions (see e.g.~\cite{Calibbi:2021pyh}):
\begin{align}
\label{eq:3lep}
\text{BR}(\ell_\alpha \to \ell_\beta \ell_\gamma \bar{\ell}_\delta) \simeq  \frac{m_{\ell_\alpha}^5}{3(16\pi)^3 \Gamma_{\ell_\alpha}} \Bigg[
&  8(1+\delta_{\beta\gamma}) \left| C^{V\,LL}_{\alpha\beta\gamma\delta} \right|^2 + 8(1+\delta_{\beta\gamma}) \left| C^{V\,RR}_{\alpha\beta\gamma\delta} \right|^2 + \\ \nonumber 
&  8 \left| C^{V\,LR}_{\alpha\beta\gamma\delta} \right|^2 + 8 \left| C^{V\,RL}_{\alpha\beta\gamma\delta} \right|^2 \Bigg]\,,
\end{align}
where we omitted the contributions of the photon dipole operator, which are loop suppressed, and neglected the masses of the final-state leptons.
Integrating out the $W^\prime$ fields, the WCs of the LFV effective operators $C^{V\,XY}_{\alpha\beta\gamma\delta}\,(\overline{\ell_\alpha} \gamma^{\mu} P_X \ell_{\beta})(\overline{\ell_\gamma} \gamma_{\mu} P_Y \ell_{\delta})$ (with $P_{X,Y}$ denoting chirality projectors) that appear in the above expression result in our case: 
\begin{equation}
C^{V\,XY}_{\alpha\beta\gamma\delta}  =  \sum_i\frac{g_F^2}{4 m_{W^{\prime}}^2}  C_{X\,\alpha\beta}^{e\,i} \, C_{Y\,\gamma\delta}^{e\,i}\,. 
\end{equation}
Based on these expressions, we find that the strongest constraint in the lepton sector follows from the upper bound $\text{BR}(\mu^{+}\to e^{+}e^{-}e^{+})<1.0\times10^{-12}$ (90\%\,CL)~\cite{SINDRUM:1985vbg}, which translates in the following lower limit on the $SU(2)_F$ breaking scale: 
\begin{equation}
\mu \to eee: \quad v_\phi \gtrsim 5.6 \times 10^4 \, \mathrm{GeV} \quad (\text{local case})\,.
\end{equation}

\paragraph{PNGBs.}
In the global $SU(2)_F$ case, heavy PNGBs give analogous contributions to the above processes.  The major difference compared to the local scenario is that the PNGBs interactions are suppressed by an additional factor $\approx m_f / m_{\pi^\prime}$, where $m_f$ is the mass of the heaviest fermion involved, as one can immediately verify performing the usual manipulation of the Lagrangian in Eq.~\eqref{eq:Lpiff} (integration by parts and insertion of the equations of motion).

The differential rates of the meson decays we are interested in can be calculated as in the ALP literature,~e.g.~\cite{Cornella:2019uxs}, giving in the limit $m_{\pi^\prime}\gg m_{M_1}$:
\begin{align}
\frac{\mathrm{d}\Gamma(M_1 \to M_2 \, \ell_\alpha \bar\ell_\beta) }{\mathrm{d}m^2_{\ell_\alpha \ell_\beta}}
\simeq &~ \frac{m_{M_1}^3 m_{\ell_\alpha}^2}{64 \pi^3\,v_\phi^4} 
\left(\left|\sum_i\frac{\hat{C}_{V\,ab}^{q\,i} \hat{C}_{V\,\alpha\beta}^{e\,i}}{m_{\pi^\prime_i}^2}\right|^2+\left|\sum_i\frac{\hat{C}_{V\,ab}^{q\,i}\hat{C}_{A\,\alpha\beta}^{e\,i}}{m_{\pi^\prime_i}^2}\right|^2\right)\times
  \nonumber\\
&~ \left[f^{M_1}_0(m^2_{\ell_\alpha \ell_\beta})\right]^2\, m^2_{\ell_\alpha \ell_\beta} 
\left(1 - \frac{m_{\ell_\alpha\ell_\beta}^2}{m^2_{M_1}} \right)\,,
\end{align}
where $m^2_{\ell_\alpha \ell_\beta}$ is the dilepton invariant mass squared and we assumed $m_{\ell_\alpha}\gg m_{\ell_\beta}$ if $\beta \neq \alpha$. 

As in the local case, the most stringent constraint on heavy PNGBs is given by kaon oscillations. Although the suppression $\sim m^2_s / m^2_{\pi^\prime}$ is much larger than the one affecting the contributions to $B_{(s)}$ oscillations ($\sim m^2_b / m^2_{\pi^\prime}$), $K^0 - \overline{K^0}$ mixing is still more relevant as a combined effect of the tighter experimental constraints and the parametric suppression of $b \to d(s)$ transitions compared to $s\to d$ processes\,---\,see the coupling matrices in Eq.~\eqref{eq:Cdmatr}.
Furthermore, in the PNGB case, only the WC of the following $LLRR$ operator is not chirality suppressed:
\begin{equation}
(\overline{s}_L \gamma^\mu d_L)(\overline{s}_R \gamma_\mu d_R): \quad \sum_i\frac{m_s^2}{4 m_{\pi^{\prime}_i}^2 v_{\phi}^2}  \hat{C}_{L\,sd}^{d\,i} \hat{C}_{R\,sd}^{d\,i}\,.
\end{equation}
Applying the bounds from $\Delta m_K$ in~\cite{Silvestrini:2018dos} to this expression, we obtain
\begin{equation}
\Delta m_K:\quad v_\phi \gtrsim 5.0 \times 10^4  \left(\frac{10\,\mathrm{GeV}}{m_{\pi^\prime}}\right)\, \mathrm{GeV}  \quad (\text{global case})\,.
\end{equation}

Moving to LFV processes, integrating out heavy PNGBs, we can write the WCs of four-lepton operators appearing in Eq.~\eqref{eq:3lep}:
\begin{equation}
C^{V\,XY}_{\alpha\beta\gamma\delta}  \simeq  \sum_i\frac{m_{\ell_\alpha}m_{\ell_\gamma}}{ m_{\pi^{\prime}_i}^2 v_\phi^2}  \hat C_{X\,\alpha\beta}^{e\,i} \, \hat C_{Y\,\gamma\delta}^{e\,i}\,,
\end{equation}
where now $X\neq Y$ and we assumed $m_{\ell_\gamma} \gg m_{\ell_\delta}$ if $\delta \neq \gamma$. As our numerical analysis will show, the additional suppression factor on the $\mu\to eee$ rate, $\sim (m_\mu m_e / m_{\pi^\prime}^2)^2$, makes this process less relevant than in the local case and, in fact, less important than the (loop-induced) radiative decay $\mu\to e \gamma$. The reason is that contributions from $\mu$ and $\tau$ loops partly overcome the lepton mass suppression.

Following \cite{Cornella:2019uxs}, we can write the contributions of the PNGBs to radiative LFV decays as
\begin{equation}
\text{BR}(\ell_{\alpha}\to\ell_{\beta}\gamma) \simeq \frac{\alpha\,m_{\ell_{\alpha}}}{2\,\Gamma_{\ell_\alpha}} \left(\left|F_{\alpha\beta}\right|^{2}+\left|G_{\alpha\beta}\right|^{2}\right)\,,
\end{equation}
where, for $\mu\to e\gamma$, retaining only diagrams with $\mu$ and $\tau$ in the loop
\begin{align}
F_{\mu e} \simeq &\frac{m_{\mu}^2}{16\pi^{2}\,v_\phi^{2}} \sum_i \hat C^{e\,i}_{A\,\mu e} \hat C^{e\,i}_{A\,\mu \mu} \,g\left(\frac{m^2_{\pi^\prime_i}}{m^2_\mu} \right)
\!+\frac{m_{\tau}^2}{32\pi^{2}\,v_\phi^{2}} \sum_i\left(
\hat C^{e\,i}_{A\,\mu \tau} \hat C^{e\,i}_{A\,\tau e} - \hat C^{e\,i}_{V\,\mu \tau} \hat C^{e\,i}_{V\,\tau e} \right)\,\widetilde{g}\left(\frac{m^2_{\pi^\prime_i}}{m^2_\tau} \right), \\
G_{\mu e} \simeq &\frac{m_{\mu}^2}{16\pi^{2}\,v_\phi^{2}} \sum_i \hat C^{e\,i}_{V\,\mu e} \hat C^{e\,i}_{A\,\mu \mu} \,g\left(\frac{m^2_{\pi^\prime_i}}{m^2_\mu} \right)
\!+ \frac{m_{\tau}^2}{32\pi^{2}\,v_\phi^{2}} \sum_i \left(
\hat C^{e\,i}_{A\,\mu \tau} \hat C^{e\,i}_{V\,\tau e} - \hat C^{e\,i}_{V\,\mu \tau} \hat C^{e\,i}_{A\,\tau e} \right)\,\widetilde{g}\left(\frac{m^2_{\pi^\prime_i}}{m^2_\tau} \right), 
\end{align}
where the loop functions $g$ and $\widetilde{g}$ are listed in the Appendix~\ref{app:loopf}. Despite the $\sim (m_\tau / m_\mu)^2$ enhancement, the second terms in the above expressions give contributions of the same order as (in fact smaller than) the first ones, because flavor-changing interactions involving the third generation are parametrically smaller due to the $SU(2)_F$ coupling structure.
Similarly, we can derive expressions for $\tau \to \ell \gamma$ ($\ell = \mu,e$) adapting the formulae in~\cite{Cornella:2019uxs} and taking into account that, in this case, $\tau$ loops are always subdominant because of the large suppression of the flavor-conserving coupling to the third generation, $\hat C^{e\,i}_{A\,\tau\tau}$.
Hence, the resulting constraints are not competitive with other limits, in particular from $\mu \to e \gamma$:
\begin{equation}
\mu \to e\gamma: \quad v_\phi \gtrsim 1.4 \times 10^5 \,\left(\frac{10\,\mathrm{GeV}}{m_{\pi^\prime}}\right)\, \mathrm{GeV}   \quad (\text{global case})\,,
\label{eq:meg}
\end{equation}
where we used the recent limit $\text{BR}(\mu^+\to e^+\gamma) < 1.5 \times 10^{-13}$ (90\%\,CL) set by the MEG~II experiment~\cite{MEGII:2025gzr}. This is the strongest flavor bound on the $SU(2)_F$ breaking scale for the global case in the heavy $m_{\pi^\prime}$ regime.

\subsection{Numerical results and discussion}

\begin{figure}[t!]
    \centering
    \begin{subfigure}[b]{0.48\textwidth}
        \centering
        \includegraphics[width=\textwidth]{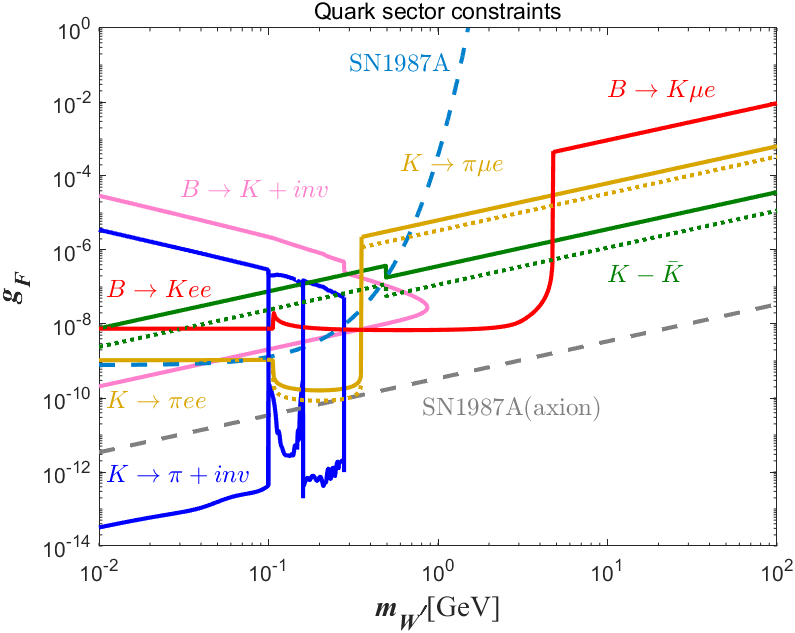}  
    \end{subfigure}
    \hfill
    \begin{subfigure}[b]{0.48\textwidth}
        \centering
        \includegraphics[width=\textwidth]{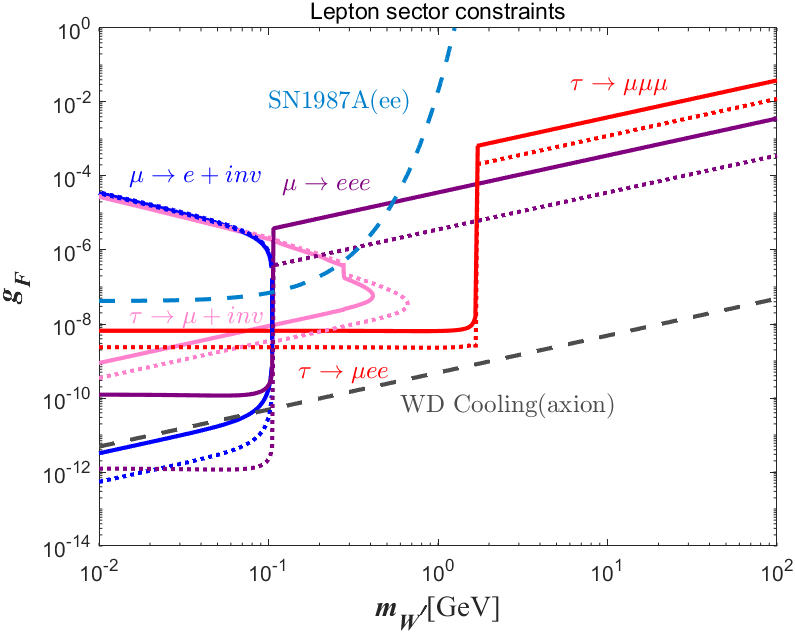} 
    \end{subfigure}   
    \caption{{\bf Local $\mathbf{SU(2)_F}$ case.} Regions of the $(m_{W^\prime},\,g_F)$ plane excluded by current experimental limits (cf.~Appendix~\ref{app:explim}) on quark sector (left) and lepton sector (right) processes mediated the $W^\prime$ gauge bosons. Regions \emph{above} or within the colored contours are excluded. Dotted lines represent expected future bounds. Dashed lines display the astrophysical limits discussed in Appendix~\ref{app:astro}. See the main text for details.} %
    \label{fig:gauge} %
\end{figure}

We can now summarize and illustrate our results, starting from the the local $SU(2)_F$ scenario. In Figure~\ref{fig:gauge}, we display constraints set by flavor processes involving a gauge boson triplet $W^\prime$ as limits on the gauge coupling $g_F$ as function of the $W^\prime$ mass. In the left panel of the figure, we show bounds from quark FCNC processes, whereas limits from purely leptonic LFV decays are displayed in the right panel. In both plots, the regions of the $(m_{W^\prime},\,g_F)$ plane lying \emph{above} or within the colored contours are excluded by the current experimental limits discussed in Appendix~\ref{app:explim}.

The figure illustrates that, if kinematically allowed, meson and lepton decays into an on-shell $W^\prime$ triplet set the most stringent constraints, as expected from the discussion in Section~\ref{sec:light-fcnc}. The recent NA62 search~\cite{NA62:2025upx} for $K\to \pi\, X_\text{inv}$ (with $X_\text{inv}$ denoting a light undetected boson) is more sensitive for tiny values of $g_F$, for which $W^\prime$ is long-lived, excluding the region within the blue contour in the left plot. Similarly, in the lepton sector, current limits on $\mu\to e \,X_\text{inv}$~\cite{TWIST:2014ymv} provide the most stringent constraint (blue contour in the right plot) for the small $m_{W^\prime}$ and small $g_F$ regime.
In particular, the $K\to \pi\, X_\text{inv}$ limit translates into the strongest bound on the $SU(2)_F$ breaking scale $v_\phi = 2 m_{W^\prime}/g_F$ for a light $W^\prime$, $v_\phi \gtrsim 6\times 10^{11}~\text{GeV}$, demonstrating once again the exceptional sensitivity of searches for exotic kaon decays to flavor-violating light new physics\,---\,see \cite{Calibbi:2016hwq,Smolkovic:2019jow,MartinCamalich:2020dfe} for other examples.

Searches for light invisible particles lose sensitivity if  $g_F$ increases \,---\,that is, the $W^\prime$ lifetime decreases\,---\,but, as Figure~\ref{fig:gauge} illustrates, various processes with the light particle decaying into a visible final state inside the detector exclude such a portion of the parameter space. The most relevant limits of this kind are provided by $K \to \pi \,e\mu/ee$ (left panel, yellow line) and $\mu\to eee$ (right panel, purple line).

For increasing values of the $W^\prime$ mass, kaon and muon decays into $W^\prime$ become kinematically forbidden, such that the latter processes give rise to much weaker constraints on the coupling $g_F$, as decay widths mediated by an off-shell $W^\prime$ scale as $g_F^4$ instead of $g_F^2$. In this regime, decays of third generation fermions into $W^\prime$ provide the most important bounds, as illustrated in Figure~\ref{fig:gauge}. Given the relatively weak experimental sensitivities compared to 1-2 transitions, limits from searches for $B$ and $\tau$ decays into invisible
bosons (pink lines) are subdominant because the $W^\prime$ lifetime is too short in the relevant $m_{W^\prime}$ range, whereas the visible counterparts of these decays set the strongest bounds: $B\to K\mu e$ (left panel, red line) in the quark sector, the LFV decays $\tau\to \mu \mu \mu$ / $\tau\to \mu ee$ (right panel, red line) in the lepton sector. The latter processes are slightly more sensitive and have better future prospects due to the expected Belle~II sensitivity to LFV $\tau$ decays~\cite{Belle-II:2018jsg,Banerjee:2022xuw}.

Finally, if the $W^\prime$ triplet is even heavier, $m_{W^\prime} > m_B$, the only relevant constraints are provided by the $W^\prime$-mediated processes discussed above in Section~\ref{sec:heavy-fcnc},\footnote{FCNC top decays are not competitive given the weak experimental limits~\cite{ParticleDataGroup:2022pth} and the relative suppression of $W^\prime$ couplings to third-generation fermions.}
the most important being provided by $K^0 - \overline{K^0}$ mixing and, again, $\mu\to eee$. The  $K^0 - \overline{K^0}$ bound (green line) is the strongest in the heavy $W^\prime$ regime\,---\,see also Eq.~\eqref{eq:DmKlocal}\,---\,and, interestingly, could increase in the near future thanks to lattice QCD calculations that are expected to improve the determination of the SM prediction for $\Delta m_K$ to 10\% level precision~\cite{Blum:2022wsz}.

\begin{figure}[t!]
    \centering
    \begin{subfigure}[b]{0.48\textwidth}
        \centering        \includegraphics[width=\textwidth]{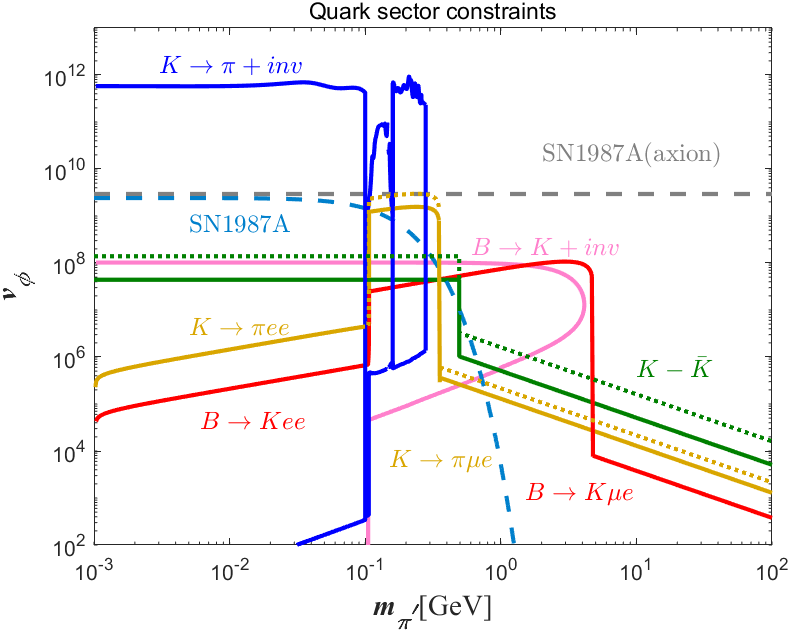} %
    \end{subfigure}
    \hfill
    \begin{subfigure}[b]{0.48\textwidth}
        \centering
        \includegraphics[width=\textwidth]{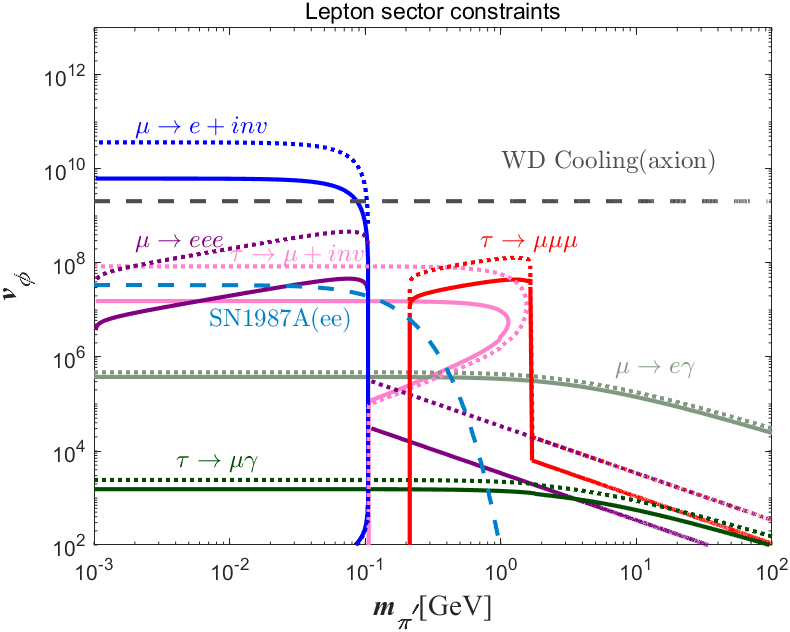} %
    \end{subfigure}
  \caption{{\bf Global $\mathbf{SU(2)_F}$ case.} Regions of the $(m_{\pi^\prime},\,v_\phi)$ plane excluded by current experimental limits (cf.~Appendix~\ref{app:explim}) on quark sector (left) and lepton sector (right) processes mediated the $\pi^\prime$ PNGBs. Regions \emph{below} or within the colored contours are excluded. Dotted lines represent expected future bounds. Dashed lines display the astrophysical limits discussed in Appendix~\ref{app:astro}. See the main text for details.}
    \label{fig:PNGB}
\end{figure}

In Figure~\ref{fig:PNGB}, we present analogous results for the global $SU(2)_F$ scenario. In this case, limits on processes mediated by or involving the $\pi^\prime$ PNGBs are displayed on the $(m_{\pi^\prime},\,v_\phi)$ plane, so that the regions \emph{below} or within the colored contours are excluded.
The Goldstone equivalence theorem guarantees that, in the $m_{\pi^\prime} \ll m_f$ limit ($m_f$ being the mass of the decaying fermion), the bounds on the flavor symmetry breaking scale from $K\to \pi\,X_\text{inv}$ ($v_\phi \gtrsim 6\times 10^{11}~\text{GeV}$) and $\mu\to e\,X_\text{inv}$ ($v_\phi \gtrsim 6\times 10^9~\text{GeV}$) are the same as in the local case with $m_{W^\prime} \ll m_f$. Nevertheless, there are also differences between the two scenarios.
In the global case, the decay width $\Gamma(\pi^\prime \to f f^{(\prime)})$ scales as $\sim m_{\pi^\prime} (m_f / v_\phi)^2$, while for the gauge bosons $\Gamma(W^\prime \to f f^{(\prime)}) \sim g_F m_{W^\prime} \sim m_{W^\prime} (m_{W^\prime} / v_\phi)^2$\,---\,see Eqs.~\eqref{eq:Wpwidth} and \eqref{eq:pipwidth}.
As a consequence, the PNGBs are substantially more long-lived, which explains why the regions excluded by  
$K\to \pi\,X_\text{inv}$ and $\mu\to e\,X_\text{inv}$ extend to much larger couplings (lower values of $v_\phi$) than in Figure~\ref{fig:gauge} and the corresponding visible decays are relatively less important. Another important difference from the local case also follows from the $\sim m_f / v_\phi$ scaling of the PNGBs couplings. As discussed above in Section~\ref{sec:heavy-fcnc}, this entails the suppression of the contributions of PNGBs to $K^0 - \overline{K^0}$ and, more drastically, $\mu \to eee$, so that the strongest limit in the heavy $\pi^\prime$ regime is given by $\mu\to e \gamma$ (Figure~\ref{fig:PNGB} right, light green line)\,---\,see also Eq.~\eqref{eq:meg}.

For comparison, Figures~\ref{fig:gauge} and~\ref{fig:PNGB} also show astrophysical bounds (cyan dashed lines) estimated as explained in Appendix~\ref{app:astro}. The most relevant constraint to the couplings of the light fields with quarks is derived from the SN1987A explosion data, while the couplings to electrons are constrained by white dwarfs~(WD) and red giants~(RG) cooling and, again, supernova explosion\,---\,the latter bound is labeled as SN1987A(ee) in the right plots. WD and RG bounds are the most stringent ($v_\phi \gtrsim \ord{10^9}~\text{GeV}$) but they do not apply to particles with $m_X \gtrsim 10-100$~keV, that is, too much above the average temperature of the star plasma to be efficiently produced. Hence, they are outside the parameter range displayed in the figures. The supernova bounds exhibit the typical Boltzmann suppression for $m_X \gtrsim 100\,\text{MeV}$ while, for simplicity, we do not calculate the loss of sensitivity in the large coupling regime when the light particles remain trapped in the star and do not contribute to its cooling. In any case, these estimates demonstrate that these astrophysical limits on light $SU(2)_F$ bosons are subdominant compared to the bounds from laboratory experiments. More important are the limits one can obtain from axiflavon emission from the star plasma, which are indicated as gray dashed lines in the figures. Recall that, in both scenarios we consider, this field is the PNGB of the global factor $U(1)_F \subset U(2)_F$ and behaves like a QCD axion acquiring a tiny mass\,---\,see Eq.~\eqref{eq:ma}\,---\,from the QCD anomaly, hence it is always light enough to be produced in the star plasma\,---\,see Section~\ref{sec:axiflavon} and Appendix~\ref{sec:axion-pheno}. The most relevant resulting constraint (from SN1987A cooling) translates to a limit $v_\phi \gtrsim 3\times 10^9~\text{GeV}$ that, however strong, does not prevent the observation of light $SU(2)_F$ bosons in kaon or muon decays, as illustrated by Figures~\ref{fig:gauge} and~\ref{fig:PNGB}. Furthermore, in the large coupling/small $v_\phi$ region also the axiflavon enters the trapping regime, which makes the other processes discussed in this section relevant tests of such a portion of the parameter space.

Interestingly, the region $v_\phi  \approx 10^{12}-10^{14}~\mathrm{GeV}$ where the axiflavon is a dark matter candidate naturally produced via the misalignment mechanism\,---\,see Section~\ref{sec:axiflavon}\,---\,is untouched by the limits discussed in this section and may be hopefully tested in the future, at least partly. In particular, this would require kaon experiments with higher sensitivity to $K\to \pi X_\text{inv}$.\footnote{From this point of view, it is unfortunate that the HIKE project~\cite{HIKE:2022qra} was recently discontinued.}

\section{Summary and outlook}
\label{sec:concl}

In this study, we have explored the implications of the $U(2)_F = SU(2)_F\times U(1)_F$ flavor model proposed in \cite{Linster:2018avp}, with a focus on the phenomenology of the bosons associated with $SU(2)_F$. Both the global and local $SU(2)_F$ scenarios entail the existence of new particles\,---\,PNGBs or gauge bosons, respectively\,---\,with distinctive interactions and phenomenological properties that imply stringent constraints on the flavor symmetry breaking scale.

By construction, in both versions of the model, two of the states belonging to the new boson triplets have unsuppressed flavor-violating couplings in the 1-2 sector. As a consequence, the existence of such particles (if light enough) is tightly constrained by searches for $K\to \pi\,X$ and $\mu\to e\,X$, as illustrated by Figures~\ref{fig:gauge} and~\ref{fig:PNGB}. The former process provides the strongest bound on the flavor symmetry breaking scale, $v_\phi \gtrsim 6\times 10^{11}~\text{GeV}$ in the light $m_X$ regime. Heavier bosons can be tested in $\tau$ and $B$ decays, which currently give limits of the order $v_\phi \gtrsim 10^{8}~\text{GeV}$. If the $SU(2)_F$ bosons are too heavy to be produced on-shell in meson or lepton decays, they can still mediate FCNC and LFV processes inducing limits in the range $v_\phi \gtrsim 10^5 - 10^6~\text{GeV}$.

Our results demonstrate that the existence of light new physics particles (which is unavoidable in the case of global continuous flavor symmetries but possible also in the local case) implies that laboratory experiments can test the flavor breaking scale, that is, the origin of fermion masses and mixing, up to ultra-high energies, in some cases even beyond the limits set by astrophysical observations.

As an outlook, we can mention a couple of directions that deserve further investigation.
For instance, it would be interesting to assess the capability of future gravitational wave (GW) observatories to test the high-energy symmetry breaking regime of this model, similarly to what done in \cite{Blasi:2024vew} for $U(1)_F$ Froggatt-Nielsen models. A possibly observable stochastic GW background (SGWB) would result from cosmic strings produced at the phase transition temperature $T \approx v_\phi$, which requires the reheating temperature $T_\text{RH}$ at the end of inflation to satisfy $T_\text{RH} > v_\phi$. This would imply that, unlike the scenario considered above, axiflavon DM is produced through misalignment after inflation and domain walls cannot be diluted by inflation. The distinctive predictions, possible issues and resulting imprints on the SGWB following from this setup deserve dedicated future studies\,---\,for studies on the SGWB from phase transitions and cosmic strings within other kinds of $SU(2)$ flavor models, see~\cite{Antusch:2025xrs,Chrysostomou:2025vrg}.
Another interesting direction is related to the cosmological role of the PNGBs $\pi^\prime$ or the gauge bosons $W^\prime$ if they are ultra-light\,---\,a natural situation in the former case. 
Even if additional DM candidates seem not to be required within this model, such ultra-light fields\,---\,if substantially produced in the early universe\,---\,could act as dark radiation or be constrained by big-bang nucleosynthesis inducing further interesting signatures or tests of the model.

\section*{Acknowledgments} 
We would like to thank Claudia Hagedorn, Tao Liu, Sida Lu, Yu-Cheng Qiu, Michael A. Schmidt, Kaifeng Zheng, and Robert Ziegler for useful discussions.
JY thanks the Department of Physics of the University of Chicago for hospitality during the completion of this work.
This research was partially supported by the National Natural Science Foundation of China (NSFC)
under grant No.~12035008 and by the Innovation and Entrepreneurship Training Program for College Students of Tianjin (No.~202410055228).


\appendix

\section*{Appendix}

\section{Loop functions}
\label{app:loopf}

The loop functions appearing in the expressions of the PNGB-mediated radiative LFV decays presented in Section~\ref{sec:pheno} read~\cite{Cornella:2019uxs}
\begin{align}
g(x)=&\frac{(x-3)x^2\log x}{x-1}-2x+1-2\sqrt{x-4}\,x^{\frac{3}{2}}\log\left(\frac{\sqrt{x-4}+\sqrt{x}}{2}\right)\,,\\
\widetilde{g}(x)=&\frac{2x^2\log x}{(x-1)^3}+\frac{1-3x}{(x-1)^2}\,.
\end{align}

\section{Experimental limits}
\label{app:explim}

Current and future expected experimental limits that we use in our analysis are listed in Table~\ref{tab:future}.

For what concerns lepton-flavor conserving meson decays (not reported in the table), in view of the long-distance effects affecting the SM predictions, we adopt the conservative limits $\mathrm{BR}(K \to \pi ee) < 10^{-7}$, $\mathrm{BR}(B \to K ee) < 10^{-7}$, which
correspond to the order of magnitude of the measured branching ratios~\cite{ParticleDataGroup:2022pth}. 
For $B\to K \mu \mu$, we compare the experimental measurements with the SM predictions for different values of $m^2_{\mu\mu}$ reported in~\cite{Datta:2022zng} and impose that the effect from $B\to K\,X(\to\mu \mu)$ lies within the combined uncertainties (at 2$\sigma$). The resulting limit is less constraining than the bound from $B\to K \mu e$ reported in the table.

\begin{table}[t!]
\centering
\begin{tabular}{llrlr}
\toprule
\textbf{Decay mode} & \multicolumn{2}{c}{\textbf{Current limit}} & \multicolumn{2}{c}{\textbf{Expected limit}} \\
\midrule
$\mu \to e\gamma$ & $1.5 \times 10^{-13}$ & \cite{MEGII:2025gzr} & $ 6\times10^{-14}$ & \cite{MEGII:2018kmf}   \\
$\mu \to ee\bar{e}$ & $1.0 \times 10^{-12}$ & \cite{SINDRUM:1985vbg} & $ 10^{-16}$ & \cite{Blondel:2013ia}   \\
$\mu \to e \,X_\text{inv}$\,$^\dagger$ & $2.6 \times 10^{-6}$ & \cite{Jodidio:1986mz} & $ 7\times10^{-8}$ & \cite{Perrevoort:2018ttp}   \\
\midrule
$\tau \to \mu \gamma$ & $4.2 \times 10^{-8}$ & \cite{Belle:2021ysv} & $6.9 \times 10^{-9}$ & \cite{Banerjee:2022xuw}    \\
$\tau \to \mu \mu\bar{\mu}$ & $1.9 \times 10^{-8}$ & \cite{Belle-II:2024sce} & $3.6 \times 10^{-10}$ & \cite{Banerjee:2022xuw}  \\
$\tau \to \mu e\bar{e}$ & $1.6 \times 10^{-8}$ & \cite{Belle-II:2025urb} & $2.9 \times 10^{-10}$ & \cite{Banerjee:2022xuw}  \\
$\tau \to \mu\, X_\text{inv}$\,$^\ddagger$ & $3.4 \times 10^{-4}$ & \cite{Belle:2025bpu} & $ 2 \times 10^{-5}$ & \cite{Calibbi:2020jvd}  \\
\midrule
$\tau \to e \gamma$ & $3.3 \times 10^{-8}$ & \cite{BaBar:2009hkt} & $9 \times 10^{-9}$ & \cite{Banerjee:2022xuw}    \\
$\tau \to ee\bar{e}$ & $2.7 \times 10^{-8}$ & \cite{Hayasaka:2010np} & $4.7 \times 10^{-10}$ & \cite{Banerjee:2022xuw}   \\
$\tau \to e \mu\bar{\mu}$ & $2.7 \times 10^{-8}$ & \cite{Hayasaka:2010np} & $ 4.5 \times 10^{-10}$ & \cite{Banerjee:2022xuw}    \\
$\tau \to e\, X_\text{inv}$\,$^\ddagger$ & $2.8 \times 10^{-4}$ & \cite{Belle:2025bpu} & $ 8 \times 10^{-6}$ & \cite{Calibbi:2020jvd}   \\
\midrule
$K^+ \to \pi^+ \mu^+ e^-$ & $1.3 \times 10^{-11}$ & \cite{Sher:2005sp} & $10^{-12}$ & \cite{Petrov:2017wza}   \\
$K^+ \to \pi^+ X_\text{inv}$\,$^\ddagger$ & $3 \times 10^{-11}$ & \cite{NA62:2025upx} &  & \\
\midrule
$B^+ \to K^+ \mu^+ e^-$ & $6.4 \times 10^{-9}$ & \cite{LHCb:2019bix} &   \\
$B^+ \to K^+ \tau^\pm \mu^\mp$ & $4.8 \times 10^{-5}$ & \cite{BaBar:2012azg} & $3.3 \times 10^{-6}$ & \cite{Belle-II:2018jsg}   \\
$B^+ \to K^+ \tau^\pm e^\mp$ & $3.0 \times 10^{-5}$ & \cite{BaBar:2012azg} & $2.1 \times 10^{-6}$ & \cite{Belle-II:2018jsg}  \\
$B^+ \to K^+ X_\text{inv}$\,$^\ddagger$  & $8 \times 10^{-6}$ & \cite{Belle-II:2023esi,Altmannshofer:2023hkn} &  & \\
\bottomrule
\end{tabular}
\caption{Current experimental limits (90\% CL) and future prospects for the flavor-violating decay modes relevant for our analysis. $^\dagger$ corresponds to a practically massless $X$ with substantial couplings to RH lepton currents; $^\ddagger$ corresponds to a massless $X$. See the text for details.}
\label{tab:future}
\end{table}

As discussed in Section~\ref{sec:light-fcnc}, for light weakly-coupled bosons, the main constraints arise from searches for an invisible particle in flavor-violating meson and lepton decays. The corresponding limits reported in Table~\ref{tab:future} show the case of a practically massless boson \mbox{$m_X \approx 0$}. In addition, the experimental limits on $\mu \to e \,X_\text{inv}$ are sensitive to the chirality of the coupling as muon experiments employ polarized beams, as discussed in details in \cite{Calibbi:2020jvd}. The displayed limit refers to a light boson with \mbox{$m_X \approx 0$} and substantial couplings with RH lepton currents (like in the case under study). In order to take into account the $m_X$ dependence for heavier bosons, we employ the average limit $\text{BR}(\mu \to e \,X_\text{inv})< 6 \times 10^{-6}$ on bosons with RH interactions obtained in~\cite{TWIST:2014ymv}. For the other processes, we adopt the exclusion curves reported in~\cite{Belle:2025bpu} for $\tau \to 
\ell \,X_\text{inv}$ (see also \cite{Belle-II:2023esi}), that on $K^+ \to \pi^+ \,X_\text{inv}$ from~\cite{NA62:2025upx} (that also includes the limit on $\pi^0\to \text{inv}$~\cite{NA62:2020fhy} in the mass gap $m_X \approx m_{\pi^0}$ where the former search is not sensitive due to the large $K^+\to \pi^+ \pi^0$ background), and finally the $2\sigma$ exclusion on $B^+ \to K^+ \,X_\text{inv}$ estimated in \cite{Altmannshofer:2023hkn} based on the Belle~II measurement of $B^+\to K^+ \nu\bar\nu$~\cite{Belle-II:2023esi}.

\revision{For certain values of the model's parameters, our light bosons could also decay inside the detector after traveling a macroscopic distance, hence giving rise to signatures commonly referred to as ``displaced vertices''. 
To the best of our knowledge, searches for displaced vertices in the decays of Table~\ref{tab:future} have never been attempted. Their viability and possible impact strongly depend on the specific process. For instance, the running Mu3e experiment is only sensitive to prompt $\mu\to ee\bar e$ decays, because the three electron tracks are required to reconstruct the primary vertex in order to tame accidental backgrounds, see e.g.~\cite{Calibbi:2017uvl}. Similarly, within $B$ factories, a secondary vertex (SV) with a significant displacement from the collision point is necessary to identify a $B$ meson decay, so that the kaon and the leptons from $ B\to K \ell\ell^\prime$ should point to such a SV. On the other hand, for a kaon decay to be visible, it is just required that the light particle $X$ decay within the decay volume (e.g.~$L \sim 100\,\mathrm{m}$ for NA62), thus it does not matter whether that occurs promptly or with some displacement. In contrast, searching for displaced $\tau$ LFV decays at Belle~II is a viable possibility that might improve the sensitivity to our light particles and requires further investigation.

In general, as we have seen in our phenomenological discussion, when limits from searches for an invisible light particle $X$ are by far the strongest, the corresponding visible decays just serve to exclude the large coupling regime. Since there is no gap between the regions excluded by searches for invisible and visible $X$ to be filled, we expect that searches for displaced vertices would not lead to increased sensitivity. Again, this is not the case of $\tau$ decays, where the visible and invisible modes are currently complementary and set competing limits, as shown by our Figures~\ref{fig:gauge} and~\ref{fig:PNGB}.
}


\section{Astrophysical limits}
\label{app:astro}

Here, we present simple estimates for the constraints on the $SU(2)_F$ bosons arising from astrophysical observations, particularly those related to star cooling. The emission of light particles from the star plasma can significantly affect stellar evolution, potentially conflicting with well-established observational data. This leads to powerful limits on the interactions of such particles with matter and radiation~\cite{Raffelt:1990yz}.

White dwarfs (WD)~\cite{Giannotti:2017hny} and red giants (RG)~\cite{Viaux:2013lha} provide stringent and complementary bounds (with the former being slightly stronger) as these stars are sensitive to additional energy loss mechanisms in case of light particles emission from electrons. Following~\cite{Calibbi:2020jvd}, the resulting constraints on the coupling to electrons of our $SU(2)_F$ bosons result
\begin{equation}
\text{WD:}\quad{v_\phi}/{\hat C_{A\,{ee}}^{e\,i}} \gtrsim 4.6 \times 10^9\,\mathrm{GeV}~(\text{global case})\,, \quad
{g_F\, C_{A\,{ee}}^{e\,i}} \lesssim 4.5 \times 10^{-13}~(\text{local case})\,. 
\end{equation}
Similarly, adapting the estimate in~\cite{Calibbi:2020jvd}, SN1987A data constrain the electron couplings as
\begin{equation}
\text{SN1987A:}\quad{v_\phi}/{\hat C_{A\,{ee}}^{e\,i}} \gtrsim 3.4 \times 10^7\,\mathrm{GeV}~(\text{global case})\,, \quad
{g_F\, C_{A\,{ee}}^{e\,i}} \lesssim 6.0 \times 10^{-11}~(\text{local case})\,. 
\end{equation}

The most relevant astrophysical constraint on couplings to quarks also comes from neutron star cooling during supernova explosions. Translating the effective coupling to nucleons as defined in~\cite{Keil:1996ju} to the case of our bosons $X$, we can write:
\begin{equation}
g_{XN} \equiv \frac{2 C_N m_N}{v_\phi}~~\text{(global case)}\,,\quad g_{XN} \equiv \frac{1}{2}g_F C_N~~\text{(local case)}\,,
\end{equation}
where $N = n, p$ denotes a neutron or a proton, $C_N$ are the dimensionless couplings whose expressions in terms of the couplings to quarks are given in \cite{Giannotti:2017hny}, and $m_N$ is the nucleon mass. The effective axion-nucleon coupling that controls energy loss is then
\begin{equation}
\bar{g}_{XN}^2 = \frac{1}{2}(g_{Xp}^2 + g_{Xn}^2) - \frac{1}{3} g_{Xp} g_{Xn}\,.
\end{equation}
Observations from SN1987A place the following limit on this effective coupling:
\begin{equation}
\bar{g}_{XN} \lesssim (0.30\text{--}0.40) \times 10^{-10}.
\end{equation}

For bosons with non-negligible mass, the energy loss due to their emission is exponentially suppressed. We follow~\cite{Raffelt:1990yz} (see also \cite{Calibbi:2020jvd}) and rescale the cooling bounds by the ratio of thermal energy densities:
\begin{equation}
R(m_X, T) \equiv \frac{\mathcal{E}_X(m_X, T)}{\mathcal{E}_X(0, T)}\,,
\end{equation}
where the energy density of a boson at temperature $T$ is given by
\begin{equation}
\mathcal{E}_X(m_X, T) = \frac{1}{2\pi^2} \int_{m_X}^\infty \frac{E^2 \sqrt{E^2 - m_X^2}}{e^{E/T} - 1} \, dE = 
\begin{cases}
\frac{\pi^2}{30} T^4, & m_X \ll T, \\
\frac{1}{(2\pi)^{3/2}} T^4 \left( \frac{m_X}{T} \right)^{5/2} e^{-m_X/T}, & m_X \gg T.
\end{cases}
\end{equation}
Since the energy loss rate scales as $v_\phi^{-2}$, the bound should be rescaled by a factor $\sqrt{R(m_X, T)}$. Due to Boltzmann suppression, cooling bounds rapidly weaken for $m_X \gtrsim 2T$, where
$T_\mathrm{RG} \approx 8.6\,\mathrm{keV}$, 
$T_\mathrm{WD} \approx 0.8\,\mathrm{keV}$, 
$T_\mathrm{NS} \approx 30\,\mathrm{MeV}$.
As a result, WD and RG constraints become ineffective for $m_X > 10-100~\text{keV}$, and we neglect them in this regime, focusing instead on SN1987A bounds.

\section{Axiflavon phenomenology}
\label{sec:axion-pheno}

Flavor and astrophysical bounds based on the axiflavon interactions presented in Section~\ref{sec:axiflavon} can be obtained as in the previous sections. Since the axiflavon is always light and long-lived, the only relevant laboratory constraints come from lepton and meson decays into an invisible particle.
Using the formulae that can be found in the literature on flavor-violating ALPs, e.g.~\cite{MartinCamalich:2020dfe,Calibbi:2020jvd}, together with the axiflavon couplings given in Section~\ref{sec:axiflavon}, we obtain:
\begin{align}
    & f_a \gtrsim 1.3\times 10^7\,\text{GeV}~~(\mu\to ea)\,,
\quad f_a \gtrsim 6.0\times 10^4\,\text{GeV}~~(\tau\to \mu a)\,, \\
& f_a \gtrsim 2.0\times 10^7\,\text{GeV}~~(K\to \pi a)\,, \quad f_a \gtrsim 1.0\times 10^6\,\text{GeV}~~(B\to K a)\,,
\end{align}
where the relation between the axion decay constant and the $U(2)_F$ breaking scale is $f_a \simeq 0.17\times  v_\phi$, see Eq.~\eqref{eq:fa}.
These limits are consistent with those calculated in \cite{Linster:2018avp} considering that we employ a different fit to the $\ord{1}$ coefficients  and a different $U(1)_F$ charge assignment in the lepton sector (both from \cite{Calibbi:2020jvd}).
As, we can see, $K \to \pi a$ and $\mu \to \pi a$ give the most stringent constraints that, however, are orders of magnitude weaker than similar limits that we obtained for the light $SU(2)_F$ bosons. This is a consequence of the $U(2)_F$ protection of the axiflavon couplings that, in particular, suppresses flavor violation in the 1-2 sector.
Therefore, more relevant are the astrophysical constraints, which read
\begin{align}
f_a \gtrsim 5\times 10^8\,\text{GeV}~~(\text{SN1987A})\,,
\quad 
f_a \gtrsim 3.5\times 10^8\,\text{GeV}~~(\text{WD})\,.
\end{align}

As discussed in Section~\ref{sec:axiflavon},
axiflavon production through the misalignment mechanism provides the correct DM abundance for $v_\phi \approx 10^{12} - 10^{14}\,\text{GeV}$ ($f_a \approx 10^{11} - 10^{13}\,\text{GeV}$).
This parameter space favored by DM can be probed by running and  upcoming haloscope experiments, see~e.g.~\cite{Irastorza:2018dyq,DiLuzio:2020wdo}. 
The relevant parameter for these searches is the axion-photon coupling~\cite{GrillidiCortona:2015jxo}:
\begin{equation}
g_{a\gamma\gamma}=\frac{1}{f_a}\frac{\alpha_\mathrm{em}}{2\pi}\left(E/N-1.92\right)\,,
\end{equation}
where the anomaly coefficients are $E=10$, $N=9/2$ in the setup we consider here, $E = 12$, $N = 9/2$ in the  $SU(5)$-compatible charge assignment originally considered in \cite{Linster:2018avp}.

\pagebreak


\bibliographystyle{JHEP}
\bibliography{U(2)refs}


\end{document}